\documentclass[aps,a4paper,reprint,prc,showpacs,showkeys,preprintnumbers,amsmath,amssymb,eqsecnum]{revtex4-1}
\usepackage{graphicx}
\usepackage{dcolumn}
\usepackage{bm}
\usepackage{url}
\usepackage{color}
\usepackage{psfrag}
\usepackage{hyperref}
\hypersetup{colorlinks}
\hypersetup{ colorlinks,
linkcolor=red,
filecolor=green,
urlcolor=red,
citecolor=blue }
\usepackage{epstopdf}
\usepackage{epsfig}
\usepackage{graphicx}
\usepackage{longtable}

\begin{document}
\preprint{Manuscript}

\title{Statistical Global Model of $\beta^-$ Halflives and r-Process 
Nucleosynthesis}\homepage{http://www.pythaim.phys.uoa.gr}\email{pythaim@phys.uoa.gr}

\author{N. J. Costiris}
\email{ncost@phys.uoa.gr}
\author{E. Mavrommatis}
\email{emavrom@phys.uoa.gr}
\affiliation{Department of Physics, Section of Nuclear  \&
Particle Physics\\
University of Athens, 15771 Athens, Greece}
\author{K. A. Gernoth}
\email{klaus.a.gernoth@manchester.ac.uk}
\affiliation{ Institute for Theoretical Physics, Johannes Kepler University Linz, 
A-4040 Linz, Austria}
\author{J.~W.~Clark}
\email{jwc@wuphys.wustl.edu}
\affiliation{McDonnell Center for the Space Sciences and
Department of Physics\\ Washington University,
St.~Louis, Missouri 63130, USA}

\date{June 24, 2013}

\begin{abstract} 
\begin{description} 
\item[Background] Reliable prediction of outcomes of stellar 
nucleosynthesis via the rapid neutron-capture process (r-process) 
continues to present major challenges to nuclear astrophysics.
Uncertainties still persist with respect to the astrophysical sites 
and the required nuclear-physics inputs.  Principally, these 
inputs involve $\beta$-decay rates of neutron-rich nuclei, to 
which both the element distribution on an r-process path 
and the time scale of the r-process are highly sensitive.
Since the majority of nuclides lying on an r-process path 
are not yet accessible experimentally, accurate forecasts
of nuclear inputs based on well-tested global models are
essential. 
\item[Purpose] 
Our objective is to apply an improved statistical global
model of $\beta^-$-decay half-life systematics~\cite{Costiris09} 
generated by machine-learning techniques to the prediction
of $\beta$ half-lives relevant to r-process nuclei. The primary 
aim of this application is to complement existing r-process-clock 
and matter-flow studies, thereby providing additional theoretical
support for the planning of future activities of the world's
network of rare-isotope laboratories.
\item[Method] The statistical model employed in this investigation 
is rooted in a fully connected, multilayer artificial neural network 
having feed-forward perceptron architecture 
$\left[{3-5-5-5-5-1\left|116\right.} 
\right]$.  The network model has been taught the systematics of 
$\beta^-$ decay based on a subset of the existing data by 
implementing the optimal Levenberg-Marquardt learning algorithm 
together with Bayesian regularization and cross-validation. 
The target domain for the modeling is the half-life systematics 
of nuclear ground states that decay 100\% by the $\beta^-$ mode, 
with cutoff at $10^{6}$ s.

\item[Results] Results are presented for nuclides situated on the 
r-ladders at $N=50$, 82, and 126 where abundances peak, as well 
as for nuclides that affect abundances between peaks or may 
be relevant to r-processes under different astrophysical 
scenarios.  The half-lives of some of the targeted neutron-rich 
nuclides have either been recently measured or will be accessible 
at rare-isotope laboratories in the relatively near future.  The 
results of our large-scale data-driven half-life calculations 
(generated by a ``theory-thin'' global statistical model) are 
compared to available experimental data, including recent 
measurements on very neutron-rich nuclei along an r-process 
path far from the valley of $\beta$ stability.  Comparison
is also made with corresponding results from traditional global 
models derived by semi-phenomenological ``theory-thick'' approaches.

\item[Conclusions]  Further evidence is presented that ``theory-thin,'' 
data-driven statistical global modeling of half-lives and other nuclear 
properties developed by machine-learning techniques can yield
reliable predictions that may serve as inputs to different 
astrophysical scenarios for the r-process.
\end{description} 
\end{abstract}


\preprint{Version 1}

\maketitle

\section{\label{sec:level-1}Introduction}

Proposed more than fifty years ago, the r-process (short for ``rapid 
neutron capture process'') is believed to be the origin of almost half
the observed elemental abundances beyond iron~\cite{Burbidge, Kappeler, 
Arnould, Kratz07, Thielemann}. The r-process abundance distributions 
(isotopic and isobaric) are normally deduced by subtracting the calculated 
s- and p-process contributions from the observed solar system abundances. 
Additionally, isotopic abundances characteristic of the early Galaxy are 
directly observed in metal-poor $\rm{Eu}$-enriched halo stars (MPEES). 
A fairly robust main r-process operating over the history of the Galaxy 
is suggested by the consistency of the MPEES abundance patterns from star 
to star with the solar system r-process abundances for the heavier 
neutron-capture elements A $\geq$130 ($\rm{Ba}$ and above).  

In spite of the numerous studies of the r-process carried out to date, there
still exist substantial uncertainties about the underlying mechanism and the 
astrophysical sites where it takes place. Since the r-process is thought 
to occur in environments featuring a very high density of free neutrons, 
potential sites are associated with core-collapse supernovae.  These include the
neutrino-driven wind in delayed explosion models, relativistic jets 
arising in failed supernovae, and magnetohydrodynamic jets from supernovae. 
Alternatively, the r-process could also occur in mergers of two neutron
stars, black-hole--neutron-star mergers, $\gamma$-ray bursts, or quark 
novae (for recent reviews, see Refs.~\cite{Thielemann, Qian}).  These 
diverse sites are inferred from corresponding models which, in addition 
to contrasting astrophysical conditions, differ in the nuclear-physics 
inputs employed in the calculations. According to the classical r-process 
model as first proposed, which is site invariant, the r-process path is 
dynamically determined by the temperature and the neutron densities. 
The equilibrium $(n,\gamma) \leftrightarrow (\gamma, n)$ fixes the path 
to very neutron-rich isotopes with one-neutron separation energies $S_n$ 
of 2--3 MeV. The nuclei within an isotonic chain at which local abundance 
peaks are called waiting-point nuclei and correspond to closed 
neutron shells (N=50, 82, 126). At such points the r-process flow pauses for 
several hundred ms (whereas the average time between two neutron 
captures is $\sim 1$ ms), since the neutron capture cross-sections are 
extremely small.  The r-process path climbs to higher $Z$ nuclei as on 
a ladder, until it reaches an isotope at which it breaks out of the magic
shell via subsequent neutron captures. At ``freeze out,'' where the 
temperature drops or the neutron density is quenched, the material 
decays back to stability via $\beta$-decay, producing the observed 
r-process abundances. The paths at $A \simeq 80$, 130, and 195 originate 
mainly from the $N=50$, 82, and 126 progenitor isotopes around 
$^{80}\rm{Zn}$, $^{130}\rm{Cd}$, and $^{195}\rm{Tm}$, respectively.  

It is evident from these considerations that various input parameters 
from nuclear physics are indispensable for the calculation of the r-process 
abundances, depending on the astrophysical model adopted. During 
the equilibrium phase the most important quantities are masses (providing 
neutron separation energies and Q-values) and the half-lives $T_{{1}/{2}}$ 
of the participating nuclei.  The masses determine the reaction path of 
the process, whereas the half-lives of the waiting-point nuclei determine 
how much material is transferred from one isotopic chain to another, 
and hence the progenitor abundances. In the ``freeze out'' phase, 
$\beta$-delayed neutron emission probabilities are needed since 
they divert the $\beta$-decay chains into neighboring nuclei and alter 
the observed abundance curve. Additionally, for nuclides above $A \geq 210$, 
neutron-capture cross-sections and alpha-decay half-lives are needed 
while for nuclides with $Z>80$, fission parameters such as barriers,
$\beta$-delayed fission probabilities, and neutron-induced fission 
cross-sections are involved. The end of the r-process, strongly 
model-dependent, is reached when the Coulomb energy becomes too 
large, most probably in the charge-mass region around $Z=94$, 
$A=270$. There, the r-process material is recycled back into 
the $A \sim 130$ region by neutron-induced, beta-delayed, and 
spontaneous fission (``fission recycling'').  

In this paper, we focus on the half-lives of nuclides involved in 
the r-process. As already indicated, knowledge of $\beta^-$ half-lives 
$T_{\beta^{-}}$ of heavy neutron-rich nuclides is essential to a 
thorough understanding of the r-process, due to their crucial role 
in determination of the time scale for matter flow and the abundances 
of heavier nuclei. Since very-neutron-rich nuclei far from the valley 
of stability participate in the r-process, the available experimental information
on relevant $T_{\beta^-}$ values has been quite limited, owing to the 
difficulty of the required measurements.  Principally, data have been 
available in those regions where the reaction path approaches the 
stable valley, specifically at or near the shell closures at $N=50$, 
82, and 126.  The pioneering experiments on isotopes along such a 
reaction path involved measurement, at ISOLDE, of the half-lives 
of the $N=50$ isotopes $^{79}{\rm Cu}$ and $^{80}\rm{Zn}$ and the 
$N=82$ isotopes $^{129}\rm{Ag}$ and $^{130}\rm{Cd}$.  All 
relevant experimental $T_{\beta^{-}}$ half-lives available up to 
November 2003 are accessible in NUBASE2003 (hereafter denoted NUBASE03)~\cite{NUBASE03}.

Experimental data on $\beta$ half-lives have subsequently been obtained 
mainly at NSCL at MSU, ISOLDE at CERN, GANIL, and RIKEN.  We refer to 
those related to the r-process in Sect.~\ref{sec:level-3}.  Currently,
experiments performed at existing radioactive-ion facilities 
(Rex-ISOLDE at CERN, FRS/ESR at GSI, RIBF at RIKEN) are under 
analysis, and further $\beta^-$ half-life measurements are scheduled. 
More advanced radioactive-beam facilities will come into operation 
in the relatively near future, with increased yields by factors up to 
1000 (FAIR at GSI, FRIB at MSU and RIBF at RIKEN~\cite{Facilities}). 
In spite of the increasing experimental activity, many neutron-rich 
nuclides contributing to the r-process will remain inaccessible 
to measurement for some time to come.  Accordingly, continued 
progress in modeling the r-process rests on reliable predictions 
from macroscopic/microscopic models of nuclear systematics based 
on fundamental nuclear theory and estimates provided by innovative 
methods of statistical inference adapted to the problem domains 
of $\beta$-decay half-lives and other relevant nuclear observables.

A number of useful approaches to traditional calculation of $\beta$-decay 
half-lives within state-of-the-art nuclear theory have been proposed 
and applied in different regions of the nuclear chart. These include 
the more phenomenological treatments, such us the gross theory (GT), 
as well as microscopic approaches based on the shell model and the 
proton-neutron quasi-particle random-phase approximation ($pn$QRPA) in 
various versions. Some of these approaches emphasize only global 
applicability, while others seek self-consistency or comprehensive 
inclusion of nuclear correlations. In the following we call attention
to the most recent theoretical calculations within these categories
and invoke their results as standards for comparison throughout the 
paper. Additionally, a brief review of a number of conventional theoretical models of 
$\beta$-decay systematics may be found in Ref.~\cite{Costiris09}.  
In the shell model (SM) calculations of Ref.~\cite{Garcia}, the 
detailed structure of the beta strength function is considered and 
results are given for nuclei at neutron number $N=82$. The size of 
the configuration space sets limits such that calculations for heavy 
nuclei are become impractical.
In the hybrid model of M\"{o}ller and co-workers~\cite{Moller}, the study 
of the nuclear ground state is based on the finite-range droplet model 
(FRDM) and a folded-Yakawa single-particle potential. The $\beta$-decay 
half-lives for the allowed Gamow-Teller transitions are obtained from a 
$pn$QRPA calculation after the addition of pairing and Gamow-Teller residual 
interactions, whereas the effect of the first-forbidden (\textit{ff}) 
transitions is included by means of the statistical gross theory 
($pn$QRPA+\textit{ff}GT). Results are derived for nuclides in the range 
$8 < Z \leq 110$ and $11 \leq N \leq 229$.  Borzov and co-workers have 
developed procedures for determining ground-state properties and 
Gamow-Teller and \textit{ff} transitions of nuclei based on a density 
functional plus continuum QRPA (DF3+CQRPA) approximation, with the 
spin-isospin effective $NN$ interaction of finite Fermi system theory 
operating in the \textit{ph} channel. This approach has been applied 
near closed neutron shells at $N=50$, 82, 126, and in the region 
``east'' of $^{208}\rm{Pb}$~\cite{Borzov, Borzov2}. These 
approaches are non relativistic. In the relativistic framework, there
is a recent $pn$QRPA calculation ($pn$RQRPA) of the Gamow-Teller 
$\beta$-decay half-lives based on a relativistic Hartree-Bogoliubov 
description of nuclear ground states that involves a density-dependent 
effective interaction DD-MEI* and includes momentum-dependent nuclear 
self-energies. This treatment has been applied in the calculation 
of $\beta$-decay half-lives $T_{\beta^{-}}$ of neutron-rich even-even 
nuclei in the $Z \sim 28$ and $Z \sim 50$ regions~\cite{Marketin}.

Although conventional theory shows continuing improvement, especially 
with respect to the evaluation of \textit{ff} transitions, the predictive 
power of these ``theory-thick'' models is rather limited far from 
stability. Most cases permitting calculation exhibit sensitivity 
to input quantities that are poorly known.  Given this situation, 
``theory-thin'' data-driven statistical modeling based on artificial 
neural networks (ANNs), and other machine-learning techniques of 
statistical inference such as support-vector machines, offers 
a potentially effective alternative for global modeling of 
$\beta^{-}$-decay lifetimes.  Such approaches have proven effective
in statistical modeling of other nuclear properties (including 
atomic masses, neutron separation energies, ground-state spin and parity, 
and branching probabilities for different decay channels), as well
as in a variety of other scientific problem domains~\cite{Clark,Gernoth}. 
The $\beta$-decay half-life is in principle determined as a mapping 
from the atomic and neutron numbers $Z$ and $N$ identifying a nuclide 
to its corresponding value $T_{\beta^-}$, and one constructs an
approximation to this mapping based on a {\it subset} of the available
data (called the training or learning set).  Statistical global
models of this kind have previously been applied to the $\beta$-decay
problem in Refs.~\cite{Mavrommatis, Clark2, Costiris09}. 

The findings of this paper are based on our most recent ANN statistical 
global model of $T_{\beta^{-}}$ systematics for the half-lives of nuclear 
ground states that decay 100\% by the $\beta^{-}$ mode (referred to
as the standard ANN model)~\cite{Costiris09}.  This model has been 
developed by implementing a number of advances in machine-learning 
algorithms. Here it will be applied in a comprehensive study of the 
$\beta^-$ half-lives of nuclides relevant to the r-process.  The 
results are compared with the available experimental data and with 
corresponding predictions from ``theory-thick'' models as cited 
above.  The ANN model exhibits good behavior in terms of several 
important performance measures. In a purely results-oriented sense, 
its predictive accuracy matches or surpasses that of traditional models 
based on microscopic nuclear theory and phenomenology.  Accordingly, 
the ANN model serves to complement and support the traditional 
``theory-thick'' models, which have the further aim of providing 
valuable insight into the underlying physics.  Section~\ref{sec:level-2} 
describes the essential features of our model.  Results for the 
half-lives of r-process nuclides are reported and discussed in 
Sect.~\ref{sec:level-3}. Finally, Sect.~\ref{sec:level-4} summarizes 
the conclusions of the present study.  
\section{\label{sec:level-2}The Model} 

\subsection{\label{sec:level-2-1}General}

Inspired by natural neural systems, an artificial neural network (ANN) 
consists of interconnected dynamical units (usually called neurons) 
that are typically arranged in a distinct layered topology and 
characterized by soft-threshold nonlinear response to weighted and 
summed inputs. The pattern of weighted connections between the units 
(along with unit biases) determines the function of the network.  The 
network employed in the present work is a multilayer {\it feed-forward}
ANN~\cite{Haykin}, thus a multilayer perceptron, whose gross architecture can be 
summarized by the notation
\begin{equation}
\left[ {I - H_1  - H_2  -  \cdots  - H_L  - O \left| {W} \right.} \right],
\label{eq:1}
\end{equation}
where $I$ is the number of inputs, $H_i$ is the number of neurons in 
the $i^{th}$ hidden layer, $O$ is the number of outputs, and $W$ is the 
total number of parameters needed to complete the specification of the 
network in terms of the weights $w_{ij}$ of the connections and the 
biases $b_i$ of the units. In global nuclear modeling, ANNs can perform
two kinds of tasks:  classification and function approximation 
(or regression), the former being a special case of the latter. In the 
latter, any nuclear observable can be viewed as a mapping from a set 
of independent variables identifying an arbitrary nuclide (here, proton
and neutron numbers $Z$ and $N$), to the corresponding dependent 
variable chosen for study (here, the $\beta{^-}$ half-life of 
the ground state of the parent nucleus).  The function so 
defined is determined by the set of parameters $\{w_{ij},b_i\}$,
whose values are derived during the learning (or training) phase 
by applying a suitable machine-learning algorithm to minimize some 
appropriate measure (called a cost function or objective function) 
of the errors made by the network in response to inputs corresponding 
to a set of examples or ``training patterns'' chosen from the mapping 
$(Z,N) \to T_{\beta^{-}}$.  A network created by such a supervised 
learning scheme is said to exhibit good generalization or prediction 
if it performs well for inputs (nuclides) belonging to a test set 
disjoint from the training set.  Statistical modeling of this sort 
inevitably involves a trade-off between closely fitting the training 
data and performing reliably in interpolation and extrapolation, 
i.e., in generalization. To enhance generalization, we employ a 
third set of patterns, called the validation set, which includes 
nuclides disjoint from both the training and test sets. 

\subsection{\label{sec:level-2-2}ANN for $T_{\beta^{-}}$ values}

In Ref.~\cite{Costiris09} we have developed a fully connected feed-forward 
artificial neural network with architecture symbolized by 
$\left[{3-5-5-5-5-1\left|116\right.} \right]$ that generates $T_{\beta^{-}}$ 
values for given nuclides. The activation functions of the processing units 
(model neurons) of the network are taken to be of hyperbolic-tangent-sigmoid 
form in the four intermediate hidden layers, a saturated linear function being 
chosen for the single neuron of the output layer.  Using existing $\beta^-$
data, this network has been taught with the Levenberg-Marquardt back-propagation
learning algorithm~\cite{Bishop}, a procedure which has the fastest convergence in 
function-approximation problems.  Our goal in training was not simply to 
attain an exact reproduction of the known half-lives, but rather to achieve an 
accurate representation of the regularities inherent in the target mapping, 
thereby promoting generalization.  To this end, we have employed two well 
established techniques for avoiding over-fitting, namely cross-validation 
and Bayesian regularization.  The method of Nguyen and Widrow~\cite{Widrow} has been used 
to initialize the free network parameters (its weights and biases) whereas 
the batch mode has been adopted for their iterative updating.  The input 
coding scheme represents $Z$ and $N$ as analog variables, scaled to fit within
the sensitive range $[-1,1]$ of the activation function assigned
to two of the three input units.  Retaining the spirit of stand-alone, 
``theory-thin'' modeling driven purely by the data, a third input unit is 
added to represent the delta-parameter, which is defined as the mean 
of the parities of $Z$ and $N$. Implementation of this parity unit facilitates
the recognition of pairing and shell effects. The base-10 logarithm of the 
$\beta^{-}$ half-life calculated by the network is represented by the activity 
of the single analog output unit. The inference process performed by the ANN 
model can therefore be epitomized in the expression:
\begin{equation}
{\rm{log}}_{10} T_{\beta} (Z,N) = {\tilde g}(Z,N, \delta) 
+ \tilde \varepsilon(Z,N, \delta),
\label{eq:2}
\end{equation}
where $\tilde g(Z,N, \delta)$ is the function that decodes the decay 
information and $\tilde \varepsilon(Z,N, \delta)$ is a random expectation 
error that represents our ignorance about the dependence of 
$T_{\beta^{-}}$ on $Z$ and $N$. This $\varepsilon$  ``noise'' term 
may reflect numerous small-scale influences on the phenomenon that 
are ``effectively chaotic,'' along with regularities missed in the 
training process.

\subsection{\label{sec:level-2-3}Data Sets}

The performance of such a statistical modeling approach necessarily depends 
on both the quantity and quality of the training data. The experimental 
$\beta$-decay half-life values used in our ANN modeling have been taken 
from the NUBASE03 compilation of nuclear 
and decay properties~\cite{NUBASE03} carried out by Audi and co-workers
at the Atomic Mass Data Center. Attention is restricted to the ground states 
of parent nuclei that decay 100\% by the $\beta^{-}$ mode (NuSet-A). We have 
considered several subsets of this specific database. The best data 
set that we have finally adopted in constructing and testing our standard ANN 
model~\cite{Costiris09} is NuSet-B, consisting of 838 nuclides ranging 
from $^{35}\rm{Na}$ to $^{247}\rm{Pu}$ and derived from NuSet-A by 
applying a cut-off at $10^6$ s. Without significant detriment to the 
prediction of $\beta^{-}$ half-lives, this creates a more homogeneous 
collection of nuclides and thereby facilitates training of the network. 
NuSet-B is further subdivided randomly into three subsets with 503 nuclides 
($\sim 60\%$) used for training the network (training or learning set) and 
167 nuclides ($\sim 20\%$) used to assess the training procedure 
(validation set), the remaining 168 nuclides ($\sim 20\%$) 
being reserved to evaluate the accuracy of prediction (test set).

\begin{table}[b]
\caption{\label{tab:results}Performance measures for the learning, 
validation, test, and whole sets, achieved by the standard ANN model 
of Ref.~\cite{Costiris09}, a network with architecture 
$\left[{3-5-5-5-5-1\left|116\right.} \right]$ trained on nuclides 
from NuSet-B.}
\begin{ruledtabular}
\begin{tabular}{ccccc}
Performance	&Learning &Validation &Test &Whole \\
Measure &Set &Set &Set &Set\\
\hline
$\sigma_{\rm rms}$	&0.53	&0.60	&0.65	&{0.57} \\
$\sigma_{\rm nms}$	&1.004	&0.995	&1.012	&{0.999} \\
$\sigma_{\rm ma}$	     &0.38	&0.41	&0.46	&{0.40} \\
\rm{R}	&0.964	&0.953	&0.947 	&{0.958} \\
\end{tabular}
\end{ruledtabular}
\label{TabIeX}
\end{table}

\subsection{\label{sec:level-2-4}Performance}

The performance of our global ANN model is first evaluated by direct 
comparison with experimental data in terms of commonly used measures 
of statistical analysis, namely the root-mean-square error 
($\sigma_{\rm rms}$), the mean absolute error ($\sigma_{\rm ma}$), and 
the normalized mean-square error ($\sigma_{\rm nms}$), as well as the 
correlation coefficient $\rm{R}$ for a linear regression analysis. 
Table \ref{TabIeX} collects results for these overall quality measures 
for the learning, validation, test sets, as well the whole set 
(NuSet-B). More problem-specific quality measures that have been 
introduced in the literature for global modeling of $\beta^{-}$ 
half-lives have also been evaluated within prescribed half-life 
domains. Specifically, these are the quantities $x_K$ and 
$\sigma_K$, $M^{(10)}$, $\sigma_{M}^{(10)}$, and $\Sigma^{(10)}$ 
defined and employed by Klapdor and co-workers~\cite{Klapdor} 
and M\"{o}ller and co-workers~\cite{Moller} respectively, and 
re-examined quantitatively for both traditional and ANN models 
in Ref.~\cite{Costiris09}.  Analysis of the performance of the 
standard network model for NuSet-B confirms a balanced behavior
of network response in all $\beta^{-}$-decay regions, with relatively 
small discrepancies between calculated and observed $\beta^{-}$ half-lives.  
Comparison of this behavior with that of ``theory-thick'' global models 
demonstrates that in a clear majority of cases, our standard statistical 
ANN model of $\beta^{-}$ half-life systematics makes predictions that 
are closer to experiment. This may be partially ascribed to its larger 
number of adjustable parameters, although it should be emphasized that 
in general having more parameters leads to better fitting but not better 
prediction. Improved performance is also shown by the standard ANN 
model~\cite{Costiris09} relative to previous ANN models~\cite{Mavrommatis, 
Clark2}.  This improvement is ascribed to its strategic advantages: 
different architecture and input encoding as well as a more advanced
training procedure. 

Clearly, the challenging goal of global modeling of nuclear properties, 
statistical or conventionally theoretical, is not only to reproduce
the experimentally known $\beta^{-}$ half-lives with some accuracy
across the known nuclear landscape, but also to demonstrate good 
extrapability (the ability to extrapolate away from existing data). 
Accordingly, further tests focusing on the performance of the ANN model
in generalization have been carried out.  These include assessment
of performance on a set of ``outlying'' nuclides, not contained in NuSet-B, 
whose half-lives have been recently measured, as well as performance 
for nuclides along particular isotopic and isotonic chains.  Also within
this spirit, we have carried out two additional modeling studies for two 
other ANN models having the same architecture as the standard model, 
i.e., $\left[{3-5-5-5-5-1\left|116\right.} \right]$~\cite{Costiris09}. 
All such tests indicate that the standard ANN model extrapolates 
satisfactorily, at least for nearby nuclei in the nuclear landscape. 
A detailed analysis of the quality of performance of this model 
has been provided in Ref.~\cite{Costiris09}.         

\section{\label{sec:level-3}Results \& Discussion}

In this section we present our results for $\beta^{-}$ half-lives 
$T_{\beta^{-}}$ of nuclides relevant to the r-process, as generated
by the ANN model. We compare these results with available experimental 
values as well as with half-lives given by the theoretical models 
highlighted in the introduction, namely those denoted
$pn$QRPA+\textit{ff}GT~\cite{Moller}, 
$pn$RQRPA+\textit{ff}GT~\cite{Marketin}, SM~\cite{Garcia}, and 
DF3+CQRPA~\cite{Borzov}.  First, we examine the results obtained for 
regions close to the waiting-points N=50, 82, and 126.  A critical 
assessment of the predictive quality of the ANN model is made by examining 
its results for isotopic chains of several nuclides situated in these 
regions, as well as for some neighboring nuclides whose $T_{\beta^{-}}$ have 
been measured {\it after} publication of the NUBASE03 database~\cite{NUBASE03} 
used in the model's construction.  Looking to the future, $T_{\beta^{-}}$ 
values are predicted (i) for nuclides that are being studied experimentally at 
FSR/ESR at GSI according to the proposals S323~\cite{Montes} and
S410~\cite{Tain}, and also (ii) for neutron-rich nuclides relevant 
to the r-process that have recently been identified at RIKEN~\cite{Ohnishi} 
and at GSI~\citep{Benlliure2}, but whose $T_{\beta^{-}}$ are yet 
to be measured. 

\subsection{\label{sec:level-3-1}Nuclei in the region of the N=50 closed 
neutron shell (around the $A \simeq 80$ peak)}

\begin{figure}[th]
%
%
\begin{psfrags}%
\psfragscanon%
%
\psfrag{s05}[b][b]{\fontsize{12}{18}\fontseries{m}\mathversion{normal}\fontshape{n}\selectfont \color[rgb]{0,0,0}\setlength{\tabcolsep}{0pt}\begin{tabular}{c}PROTONS (Z)\end{tabular}}%
\psfrag{s06}[t][t]{\fontsize{12}{18}\fontseries{m}\mathversion{normal}\fontshape{n}\selectfont \color[rgb]{0,0,0}\setlength{\tabcolsep}{0pt}\begin{tabular}{c}NEUTRONS (N)\end{tabular}}%
\psfrag{s09}[b][b]{\fontsize{16}{24}\fontseries{m}\mathversion{normal}\fontshape{n}\selectfont \color[rgb]{0,0,0}\setlength{\tabcolsep}{0pt}\begin{tabular}{c}Chart of nuclides up to N = 50\end{tabular}}%
\psfrag{s10}[][]{\fontsize{10}{15}\fontseries{m}\mathversion{normal}\fontshape{n}\selectfont \color[rgb]{0,0,0}\setlength{\tabcolsep}{0pt}\begin{tabular}{c} \end{tabular}}%
\psfrag{s11}[][]{\fontsize{10}{15}\fontseries{m}\mathversion{normal}\fontshape{n}\selectfont \color[rgb]{0,0,0}\setlength{\tabcolsep}{0pt}\begin{tabular}{c} \end{tabular}}%
\psfrag{s12}[l][l]{\fontsize{11}{16}\fontseries{m}\mathversion{normal}\fontshape{n}\selectfont \color[rgb]{0,0,0}Unknown}%
\psfrag{s13}[l][l]{\fontsize{11}{16}\fontseries{m}\mathversion{normal}\fontshape{n}\selectfont \color[rgb]{0,0,0}NUBASE03~\cite{NUBASE03}}%
\psfrag{s14}[l][l]{\fontsize{11}{16}\fontseries{m}\mathversion{normal}\fontshape{n}\selectfont \color[rgb]{0,0,0}New Exp.~\cite{Hosmer,Hosmer2010,Daugas}}%
\psfrag{s15}[l][l]{\fontsize{11}{16}\fontseries{m}\mathversion{normal}\fontshape{n}\selectfont \color[rgb]{0,0,0}Stable}%
\psfrag{s16}[l][l]{\fontsize{11}{16}\fontseries{m}\mathversion{normal}\fontshape{n}\selectfont \color[rgb]{0,0,0}Unknown}%
%
\fontsize{12}{18}\fontseries{m}\mathversion{normal}%
\fontshape{n}\selectfont%
%
\psfrag{x01}[t][t]{$38$}%
\psfrag{x02}[t][t]{$40$}%
\psfrag{x03}[t][t]{$42$}%
\psfrag{x04}[t][t]{$44$}%
\psfrag{x05}[t][t]{$46$}%
\psfrag{x06}[t][t]{$48$}%
\psfrag{x07}[t][t]{$50$}%
%
\psfrag{v01}[r][r]{}%
\psfrag{v02}[r][r]{$_{22}\rm{Ti}$}%
\psfrag{v03}[r][r]{$_{23}\rm{V}$}%
\psfrag{v04}[r][r]{$_{24}\rm{Cr}$}%
\psfrag{v05}[r][r]{$_{25}\rm{Mn}$}%
\psfrag{v06}[r][r]{$_{26}\rm{Fe}$}%
\psfrag{v07}[r][r]{$_{27}\rm{Co}$}%
\psfrag{v08}[r][r]{$_{28}\rm{Ni}$}%
\psfrag{v09}[r][r]{$_{29}\rm{Cu}$}%
\psfrag{v10}[r][r]{$_{30}\rm{Zn}$}%
\psfrag{v11}[r][r]{$_{31}\rm{Ga}$}%
\psfrag{v12}[r][r]{}%
%
\resizebox{9cm}{!}{\includegraphics{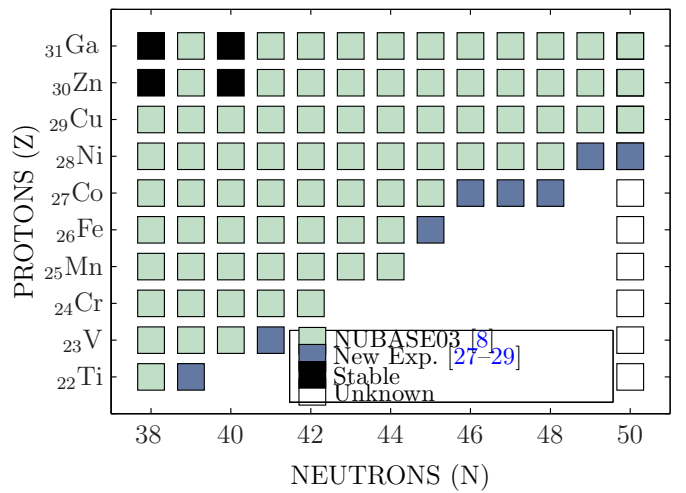}}%
\end{psfrags}%
%

\caption{\label{fig:n50chart}(Color online) Cluster plot of the chart 
of nuclides up to the $N=50$ closed shell. Black boxes indicate stable 
nuclides. Gray (green) boxes indicate the nuclides with measured 
$\beta^{-}$-decay half-lives $T_{\beta^{-}}$ reported in 
NUBASE03~\cite{NUBASE03}, while dark gray (blue) boxes identify those with 
newly measured $T_{\beta^{-}}$ beyond NUBASE03~\cite{Hosmer,Hosmer2010,
Daugas}. Empty boxes denote nuclides at the $N=50$ closed shell for which no 
experimental $\beta^{-}$ half-lives are available. Theoretical values 
for $T_{\beta^{-}}$ from the standard ANN model, as well as from 
conventional nuclear models, are plotted in Fig.~\ref{fig:n502}.}
\end{figure}

\begin{figure}[th]
%
%
\begin{psfrags}%
\psfragscanon%
%
\psfrag{s05}[b][b]{\fontsize{12}{18}\fontseries{m}\mathversion{normal}\fontshape{n}\selectfont \color[rgb]{0,0,0}\setlength{\tabcolsep}{0pt}\begin{tabular}{c}$T_{{\beta}^{-}} (ms)$\end{tabular}}%
\psfrag{s06}[t][t]{\fontsize{12}{18}\fontseries{m}\mathversion{normal}\fontshape{n}\selectfont \color[rgb]{0,0,0}\setlength{\tabcolsep}{0pt}\begin{tabular}{c}MASS NUMBER\end{tabular}}%
\psfrag{s07}[b][b]{\fontsize{16}{24}\fontseries{m}\mathversion{normal}\fontshape{n}\selectfont \color[rgb]{0,0,0}\setlength{\tabcolsep}{0pt}\begin{tabular}{c}N = 50\end{tabular}}%
\psfrag{s10}[][]{\fontsize{10}{15}\fontseries{m}\mathversion{normal}\fontshape{n}\selectfont \color[rgb]{0,0,0}\setlength{\tabcolsep}{0pt}\begin{tabular}{c} \end{tabular}}%
\psfrag{s11}[][]{\fontsize{10}{15}\fontseries{m}\mathversion{normal}\fontshape{n}\selectfont \color[rgb]{0,0,0}\setlength{\tabcolsep}{0pt}\begin{tabular}{c} \end{tabular}}%
\psfrag{s12}[l][l]{\fontsize{11}{16}\fontseries{m}\mathversion{normal}\fontshape{n}\selectfont \color[rgb]{0,0,0}$pn$RQRPA+\textit{ff}~\cite{Marketin}}%
\psfrag{s13}[l][l]{\fontsize{11}{16}\fontseries{m}\mathversion{normal}\fontshape{n}\selectfont \color[rgb]{0,0,0}NUBASE03~\cite{NUBASE03}}%
\psfrag{s14}[l][l]{\fontsize{11}{16}\fontseries{m}\mathversion{normal}\fontshape{n}\selectfont \color[rgb]{0,0,0}New Exp.~\cite{Hosmer}}%
\psfrag{s15}[l][l]{\fontsize{11}{16}\fontseries{m}\mathversion{normal}\fontshape{n}\selectfont \color[rgb]{0,0,0}ANN~\cite{Costiris09}}%
\psfrag{s16}[l][l]{\fontsize{11}{16}\fontseries{m}\mathversion{normal}\fontshape{n}\selectfont \color[rgb]{0,0,0}$pn$QRPA+\textit{ff}GT~\cite{Moller}}%
\psfrag{s17}[l][l]{\fontsize{11}{16}\fontseries{m}\mathversion{normal}\fontshape{n}\selectfont \color[rgb]{0,0,0}DF3+CQRPA~\cite{Borzov}}%
\psfrag{s18}[l][l]{\fontsize{11}{16}\fontseries{m}\mathversion{normal}\fontshape{n}\selectfont \color[rgb]{0,0,0}$pn$RQRPA+\textit{ff}~\cite{Marketin}}%
%
\fontsize{12}{18}\fontseries{m}\mathversion{normal}%
\fontshape{n}\selectfont%
%
\psfrag{x01}[t][t]{$72$}%
\psfrag{x02}[t][t]{$74$}%
\psfrag{x03}[t][t]{$76$}%
\psfrag{x04}[t][t]{$78$}%
\psfrag{x05}[t][t]{$80$}%
\psfrag{x06}[t][t]{$82$}%
\psfrag{x07}[t][t]{$84$}%
%
\psfrag{v01}[r][r]{$10^{0}$}%
\psfrag{v02}[r][r]{$10^{1}$}%
\psfrag{v03}[r][r]{$10^{2}$}%
\psfrag{v04}[r][r]{$10^{3}$}%
\psfrag{v05}[r][r]{$10^{4}$}%
\psfrag{v06}[r][r]{$10^{5}$}%
\psfrag{v07}[r][r]{$10^{6}$}%
\psfrag{v08}[r][r]{$10^{7}$}%
%
\resizebox{9cm}{!}{\includegraphics{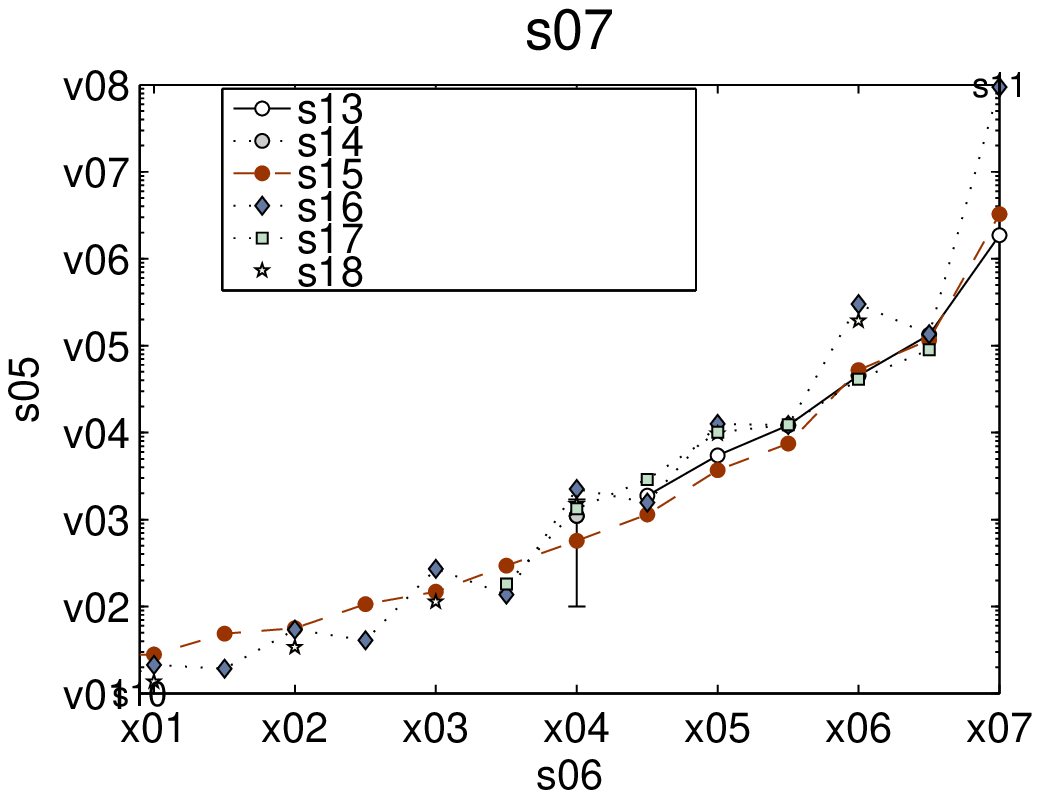}}%
\end{psfrags}%
%

\caption{\label{fig:n502}(Color online) Beta-decay half-lives 
$T_{\beta^{-}}$ produced by the ANN model for the r-ladder isotonic chain 
at $N=50$ (from $Z=22$ up to $Z=34$) in comparison with experimental 
values~\cite{NUBASE03, Hosmer}, and with corresponding results from 
$pn$QRPA+\textit{ff}GT~\cite{Moller}, DF3+CQRPA~\cite{Borzov} and 
$pn$-RQRPA+\textit{ff}~\cite{Marketin} calculations.}
\end{figure}

\begin{figure} [h]
%
%
\begin{psfrags}%
\psfragscanon%
%
\psfrag{s05}[b][b]{\fontsize{16}{24}\fontseries{m}\mathversion{normal}\fontshape{n}\selectfont \color[rgb]{0,0,0}\setlength{\tabcolsep}{0pt}\begin{tabular}{c}$_{27}\rm{Co}$\end{tabular}}%
\psfrag{s07}[t][t]{\fontsize{12}{18}\fontseries{m}\mathversion{normal}\fontshape{n}\selectfont \color[rgb]{0,0,0}\setlength{\tabcolsep}{0pt}\begin{tabular}{c}MASS NUMBER\end{tabular}}%
\psfrag{s08}[b][b]{\fontsize{12}{18}\fontseries{m}\mathversion{normal}\fontshape{n}\selectfont \color[rgb]{0,0,0}\setlength{\tabcolsep}{0pt}\begin{tabular}{c}$T_{{\beta}^{-}} (ms)$\end{tabular}}%
\psfrag{s10}[][]{\fontsize{10}{15}\fontseries{m}\mathversion{normal}\fontshape{n}\selectfont \color[rgb]{0,0,0}\setlength{\tabcolsep}{0pt}\begin{tabular}{c} \end{tabular}}%
\psfrag{s11}[][]{\fontsize{10}{15}\fontseries{m}\mathversion{normal}\fontshape{n}\selectfont \color[rgb]{0,0,0}\setlength{\tabcolsep}{0pt}\begin{tabular}{c} \end{tabular}}%
\psfrag{s12}[l][l]{\fontsize{11}{16}\fontseries{m}\mathversion{normal}\fontshape{n}\selectfont \color[rgb]{0,0,0}DF3+CQRPA~\cite{Borzov}}%
\psfrag{s13}[l][l]{\fontsize{11}{16}\fontseries{m}\mathversion{normal}\fontshape{n}\selectfont \color[rgb]{0,0,0}NUBASE03~\cite{NUBASE03}}%
\psfrag{s14}[l][l]{\fontsize{11}{16}\fontseries{m}\mathversion{normal}\fontshape{n}\selectfont \color[rgb]{0,0,0}New Exp.~\cite{Daugas, Hosmer2010}}%
\psfrag{s15}[l][l]{\fontsize{11}{16}\fontseries{m}\mathversion{normal}\fontshape{n}\selectfont \color[rgb]{0,0,0}ANN~\cite{Costiris09}}%
\psfrag{s16}[l][l]{\fontsize{11}{16}\fontseries{m}\mathversion{normal}\fontshape{n}\selectfont \color[rgb]{0,0,0}$pn$QRPA+\textit{ff}GT~\cite{Moller}}%
\psfrag{s17}[l][l]{\fontsize{11}{16}\fontseries{m}\mathversion{normal}\fontshape{n}\selectfont \color[rgb]{0,0,0}DF3+CQRPA~\cite{Borzov}}%
%
\fontsize{12}{18}\fontseries{m}\mathversion{normal}%
\fontshape{n}\selectfont%
%
\psfrag{x01}[t][t]{$65$}%
\psfrag{x02}[t][t]{$70$}%
\psfrag{x03}[t][t]{$75$}%
\psfrag{x04}[t][t]{$80$}%
%
\psfrag{v01}[r][r]{$10^{0}$}%
\psfrag{v02}[r][r]{$10^{1}$}%
\psfrag{v03}[r][r]{$10^{2}$}%
\psfrag{v04}[r][r]{$10^{3}$}%
\psfrag{v05}[r][r]{$10^{4}$}%
\psfrag{v06}[r][r]{$10^{5}$}%
%
\resizebox{9cm}{!}{\includegraphics{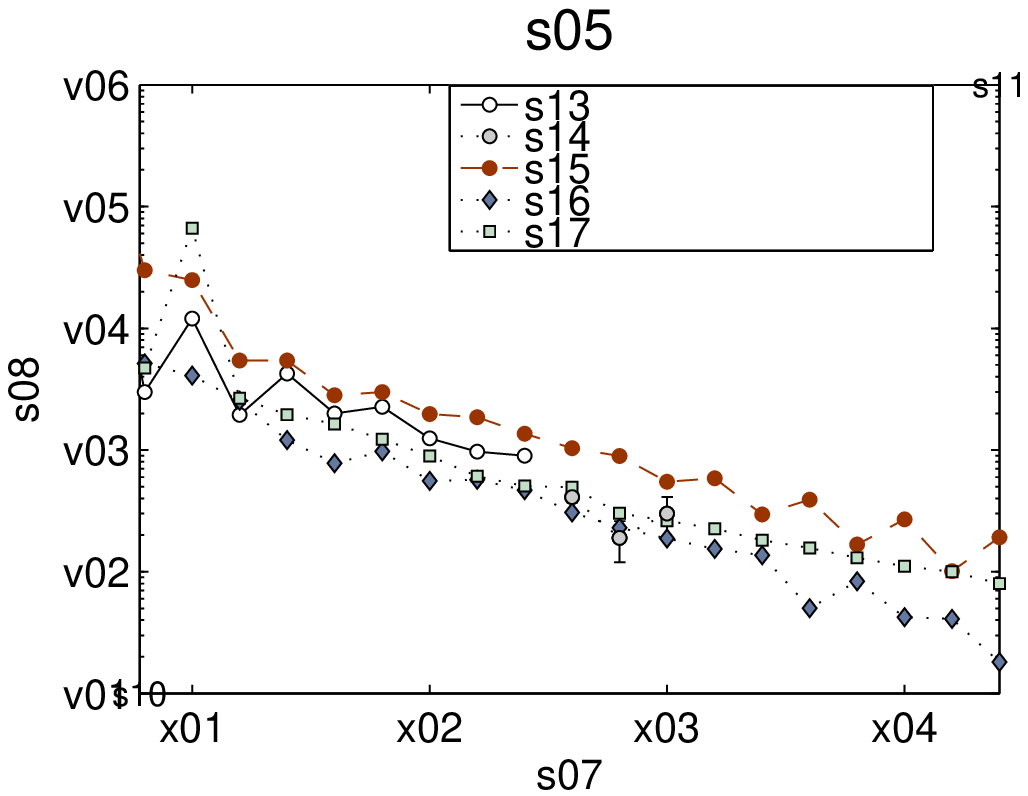}}%
\end{psfrags}%
%

\caption{\label{fig:co27}(Color online) Beta-decay half-lives 
$T_{\beta^{-}}$ produced by the ANN model for the isotopic chain of 
$_{27}{\rm Co}$ (from $N=37$ up to $N=55$) in comparison with experimental 
values~\cite{NUBASE03, Daugas, Hosmer2010} and results from   
$pn$QRPA+\textit{ff}GT~\cite{Moller} and DF3+CQRPA~\cite{Borzov} calculations.}
\end{figure}

\begin{figure} [h]
%
%
\begin{psfrags}%
\psfragscanon%
%
\psfrag{s05}[b][b]{\fontsize{16}{24}\fontseries{m}\mathversion{normal}\fontshape{n}\selectfont \color[rgb]{0,0,0}\setlength{\tabcolsep}{0pt}\begin{tabular}{c}$_{28}\rm{Ni}$\end{tabular}}%
\psfrag{s07}[t][t]{\fontsize{12}{18}\fontseries{m}\mathversion{normal}\fontshape{n}\selectfont \color[rgb]{0,0,0}\setlength{\tabcolsep}{0pt}\begin{tabular}{c}MASS NUMBER\end{tabular}}%
\psfrag{s08}[b][b]{\fontsize{12}{18}\fontseries{m}\mathversion{normal}\fontshape{n}\selectfont \color[rgb]{0,0,0}\setlength{\tabcolsep}{0pt}\begin{tabular}{c}$T_{{\beta}^{-}} (ms)$\end{tabular}}%
\psfrag{s10}[][]{\fontsize{10}{15}\fontseries{m}\mathversion{normal}\fontshape{n}\selectfont \color[rgb]{0,0,0}\setlength{\tabcolsep}{0pt}\begin{tabular}{c} \end{tabular}}%
\psfrag{s11}[][]{\fontsize{10}{15}\fontseries{m}\mathversion{normal}\fontshape{n}\selectfont \color[rgb]{0,0,0}\setlength{\tabcolsep}{0pt}\begin{tabular}{c} \end{tabular}}%
\psfrag{s12}[l][l]{\fontsize{11}{16}\fontseries{m}\mathversion{normal}\fontshape{n}\selectfont \color[rgb]{0,0,0}$pn$RQRPA+\textit{ff}~\cite{Marketin}}%
\psfrag{s13}[l][l]{\fontsize{11}{16}\fontseries{m}\mathversion{normal}\fontshape{n}\selectfont \color[rgb]{0,0,0}NUBASE03~\cite{NUBASE03}}%
\psfrag{s14}[l][l]{\fontsize{11}{16}\fontseries{m}\mathversion{normal}\fontshape{n}\selectfont \color[rgb]{0,0,0}New Exp.~\cite{Hosmer}}%
\psfrag{s15}[l][l]{\fontsize{11}{16}\fontseries{m}\mathversion{normal}\fontshape{n}\selectfont \color[rgb]{0,0,0}ANN~\cite{Costiris09}}%
\psfrag{s16}[l][l]{\fontsize{11}{16}\fontseries{m}\mathversion{normal}\fontshape{n}\selectfont \color[rgb]{0,0,0}$pn$QRPA+\textit{ff}GT~\cite{Moller}}%
\psfrag{s17}[l][l]{\fontsize{11}{16}\fontseries{m}\mathversion{normal}\fontshape{n}\selectfont \color[rgb]{0,0,0}DF3+CQRPA~\cite{Borzov}}%
\psfrag{s18}[l][l]{\fontsize{11}{16}\fontseries{m}\mathversion{normal}\fontshape{n}\selectfont \color[rgb]{0,0,0}$pn$RQRPA+\textit{ff}~\cite{Marketin}}%
%
\fontsize{12}{18}\fontseries{m}\mathversion{normal}%
\fontshape{n}\selectfont%
%
\psfrag{x01}[t][t]{$70$}%
\psfrag{x02}[t][t]{$75$}%
\psfrag{x03}[t][t]{$80$}%
\psfrag{x04}[t][t]{$85$}%
%
\psfrag{v01}[r][r]{$10^{0}$}%
\psfrag{v02}[r][r]{$10^{1}$}%
\psfrag{v03}[r][r]{$10^{2}$}%
\psfrag{v04}[r][r]{$10^{3}$}%
\psfrag{v05}[r][r]{$10^{4}$}%
\psfrag{v06}[r][r]{$10^{5}$}%
\psfrag{v07}[r][r]{$10^{6}$}%
%
\resizebox{9cm}{!}{\includegraphics{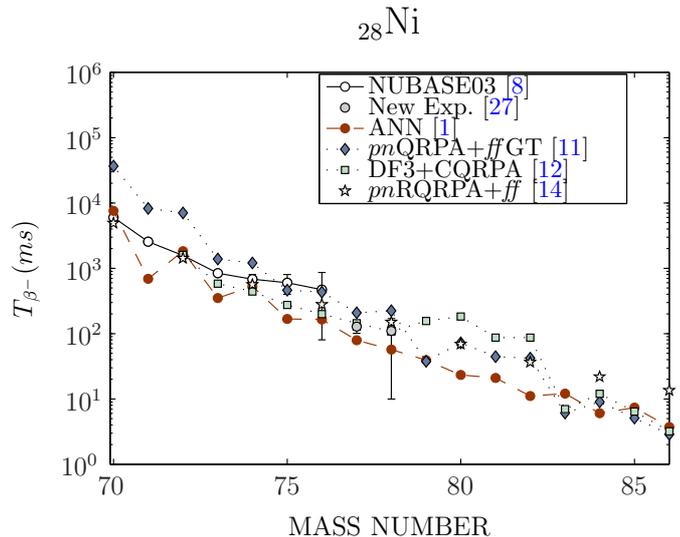}}%
\end{psfrags}%
%

\caption{\label{fig:ni28}(Color online) Beta-decay half-lives 
$T_{\beta^{-}}$ produced by the ANN model for the isotopic chain of 
$_{28}{\rm Ni}$ (from $N=42$ up to $N=58$) in comparison with 
experimental values~\cite{NUBASE03, Hosmer}, and with corresponding 
results from $pn$QRPA+\textit{ff}GT~\cite{Moller}, DF3+CQRPA~\cite{Borzov} 
and $pn$-RQRPA+\textit{ff}~\cite{Marketin} (the last only for 
even-even nuclei) calculations.}
\end{figure}

Neutron-rich nuclides in the region of the $N=50$ closed neutron shell play 
an important role in r-process nucleosynthesis. In the classical 
r-process model, the $N=50$ nuclides act as waiting-points and determine 
the formation and shape of the $A \simeq 80$ abundance peak with progenitor 
nuclide $^{80}Zn$. In the newer site-dependent r-process models, involving
a neutrino-driven wind and a high-entropy wind and extensions thereof, the 
r-process begins at lighter nuclei, but the half-lives of waiting-point 
nuclides are a direct input and set the r-process time scale through 
the $N=50$ bottleneck towards heavier nuclei.

A chart of nuclides up to the $N=50$ closed shell with known 
experimental $T_{\beta^{-}}$ is presented in Fig~\ref{fig:n50chart}. 
Nuclides whose $T_{\beta^{-}}$ value is included in 
NUBASE03~\cite{NUBASE03}  are shown with gray (green) boxes. 
Nuclides with $T_{\beta^{-}}$ values ({\it not} included in NUBASE03) 
but recently measured at NSCL~\cite{Hosmer, Hosmer2010} 
($^{75}{\rm{Co}}$, $^{77,78}{\rm{Ni}}$, $^{80}{\rm{Cu}}$) and at 
GANIL~\cite{Daugas} ($^{61}{\rm{Ti}}$, $^{64}{\rm{V}}$, $^{71}{\rm{Fe}}$, 
$^{73,74}{\rm{Co}}$) are labeled in dark gray (blue). 
The half-lives for $^{73,74}{\rm{Co}}$ have also been measured at 
NSCL~\cite{Hosmer}, but we adopt here the most recently measured 
GANIL values.

These ``beyond NUBASE03'' values along with the corresponding 
predictions from our ANN model and those available from 
$pn$QRPA+\textit{ff}GT~\cite{Moller} and DF3+CQRPA\cite{Borzov} 
calculations are displayed in Table~\ref{TabIeN126}. In 
these cases, the ANN is making true predictions rather than 
regurgitating approximate fits, since none of the nuclei 
involved were present in the training set or validation set.

One of the pivotal measurements in this isotopic region is the accurate 
determination of $T_{\beta^{-}}$ for the doubly magic waiting-point 
nucleus $^{78}{\rm Ni}$, carried out by an NSCL team at Michigan State 
University (MSU)~\cite{Hosmer}.  This quantity is among those needed 
to disentangle the various contributions from neutron-capture processes 
in different astrophysical sites.  It is of special importance since 
the time-scale for building heavy elements beyond $N=50$ is set 
by the sum of the lifetimes of $^{78}{\rm{Ni}}$ and $^{79}{\rm{Cu}}$, and 
the shorter half-life found implies an acceleration of the r-process. 
It is seen from Fig.~\ref{fig:n50chart} that all important $N=50$ 
waiting-pointing nuclei ($^{78}{\rm{Ni}}$,$^{79}{\rm{Cu}}$, $^{80}{\rm{Zn}}$) 
already have known experimental values 
for $T_{\beta^{-}}$, 
whereas this is not the case for the lighter nuclides.  
Fig.~\ref{fig:n502} shows the predictions from the ANN model for 
the isotonic chain at $N=50$ (from $Z=22$ up to $Z=34$) 
and compares them with the available experimental values and 
with results from the three theoretical models 
$pn$QRPA+\textit{ff}GT~\cite{Moller}, 
DF3+CQRPA~\cite{Borzov}, and (for even-even nuclei only) 
$pn$-RQRPA+\textit{ff}~\cite{Marketin}.  Results provided by the 
ANN model are close to the experimental values and to the results 
of $pn$QRPA+\textit{ff}GT~\cite{Moller} and DF3+CQRPA~\cite{Borzov} 
calculations. The $T_{\beta^{-}}$ values obtained by M\"{o}ller 
et al.~\cite{Moller} are larger than those given by the ANN model 
for $Z\geq 28$ and smaller for $Z<28$, and they show more pronounced 
pairing behavior. The performance of our model in this region 
can also be evaluated by studying the isotopic chains of $\rm{Co}$ 
and $\rm{Ni}$ (see Figs.~\ref{fig:co27} and \ref{fig:ni28}). 
The ANN model gives half-lives quite close to the experimental values
and in most cases shows good agreement with results of the 
DF3+CQRPA~\cite{Borzov} calculations.

\subsection{\label{sec:level-3-2}Nuclei in the region of the $N=82$ 
closed neutron shell (around the $A \simeq 130$ peak)} 

\begin{figure} 
%
%
\begin{psfrags}%
\psfragscanon%
%
\psfrag{s05}[b][b]{\fontsize{12}{18}\fontseries{m}\mathversion{normal}\fontshape{n}\selectfont \color[rgb]{0,0,0}\setlength{\tabcolsep}{0pt}\begin{tabular}{c}PROTONS (Z)\end{tabular}}%
\psfrag{s06}[t][t]{\fontsize{12}{18}\fontseries{m}\mathversion{normal}\fontshape{n}\selectfont \color[rgb]{0,0,0}\setlength{\tabcolsep}{0pt}\begin{tabular}{c}NEUTRONS (N)\end{tabular}}%
\psfrag{s09}[b][b]{\fontsize{16}{24}\fontseries{m}\mathversion{normal}\fontshape{n}\selectfont \color[rgb]{0,0,0}\setlength{\tabcolsep}{0pt}\begin{tabular}{c}Chart of nuclides up to N = 82\end{tabular}}%
\psfrag{s10}[][]{\fontsize{10}{15}\fontseries{m}\mathversion{normal}\fontshape{n}\selectfont \color[rgb]{0,0,0}\setlength{\tabcolsep}{0pt}\begin{tabular}{c} \end{tabular}}%
\psfrag{s11}[][]{\fontsize{10}{15}\fontseries{m}\mathversion{normal}\fontshape{n}\selectfont \color[rgb]{0,0,0}\setlength{\tabcolsep}{0pt}\begin{tabular}{c} \end{tabular}}%
\psfrag{s12}[l][l]{\fontsize{11}{16}\fontseries{m}\mathversion{normal}\fontshape{n}\selectfont \color[rgb]{0,0,0}Unknown}%
\psfrag{s13}[l][l]{\fontsize{11}{16}\fontseries{m}\mathversion{normal}\fontshape{n}\selectfont \color[rgb]{0,0,0}NUBASE03~\cite{NUBASE03}}%
\psfrag{s14}[l][l]{\fontsize{11}{16}\fontseries{m}\mathversion{normal}\fontshape{n}\selectfont \color[rgb]{0,0,0}New Exp.~\cite{Nishimura,Montes2,Pereira2009}}%
\psfrag{s15}[l][l]{\fontsize{11}{16}\fontseries{m}\mathversion{normal}\fontshape{n}\selectfont \color[rgb]{0,0,0}Stable}%
\psfrag{s16}[l][l]{\fontsize{11}{16}\fontseries{m}\mathversion{normal}\fontshape{n}\selectfont \color[rgb]{0,0,0}Unknown}%
%
\fontsize{13}{19.5}\fontseries{m}\mathversion{normal}%
\fontshape{n}\selectfont%
%
\psfrag{x01}[t][t]{$64$}%
\psfrag{x02}[t][t]{$66$}%
\psfrag{x03}[t][t]{$68$}%
\psfrag{x04}[t][t]{$70$}%
\psfrag{x05}[t][t]{$72$}%
\psfrag{x06}[t][t]{$74$}%
\psfrag{x07}[t][t]{$76$}%
\psfrag{x08}[t][t]{$78$}%
\psfrag{x09}[t][t]{$80$}%
\psfrag{x10}[t][t]{$82$}%
%
\psfrag{v01}[r][r]{}%
\psfrag{v02}[r][r]{$_{38}\rm{Sr}$}%
\psfrag{v03}[r][r]{$_{39}\rm{Y}$}%
\psfrag{v04}[r][r]{$_{40}\rm{Zr}$}%
\psfrag{v05}[r][r]{$_{41}\rm{Nb}$}%
\psfrag{v06}[r][r]{$_{42}\rm{Mo}$}%
\psfrag{v07}[r][r]{$_{43}\rm{Tc}$}%
\psfrag{v08}[r][r]{$_{44}\rm{Ru}$}%
\psfrag{v09}[r][r]{$_{45}\rm{Rh}$}%
\psfrag{v10}[r][r]{$_{46}\rm{Pd}$}%
\psfrag{v11}[r][r]{$_{47}\rm{Ag}$}%
\psfrag{v12}[r][r]{$_{48}\rm{Cd}$}%
\psfrag{v13}[r][r]{$_{49}\rm{In}$}%
\psfrag{v14}[r][r]{$_{50}\rm{Sn}$}%
\psfrag{v15}[r][r]{}%
%
\resizebox{9cm}{!}{\includegraphics{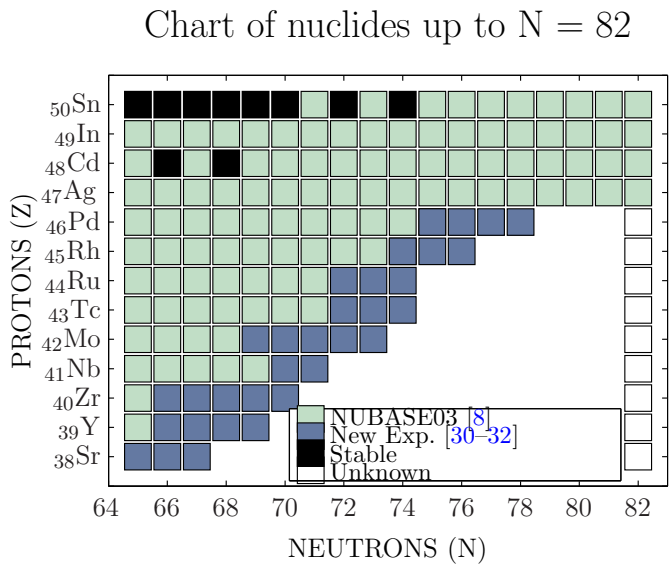}}%
\end{psfrags}%
%

\caption{\label{fig:n82chart}(Color online) Cluster plot of the chart 
of nuclides up to the $N=82$ closed shell. Black boxes indicate stable 
nuclides.  Gray (green) boxes indicate the nuclides with 
measured $\beta^{-}$-decay half-lives $T_{\beta^{-}}$ reported in 
NUBASE03~\cite{NUBASE03}, while dark gray (blue) boxes indicate those with 
newly measured $T_{\beta^{-}}$ beyond NUBASE03~\cite{Nishimura, Montes2, 
Pereira2009}. Empty boxes denote nuclides at the $N=82$ closed shell for 
which no experimental $T_{\beta^{-}}$ values are available. Predicted
values for the latter from our ANN model as well as calculated
results from three theory-thick models are given in 
Fig.~\ref{fig:n822}.}
\end{figure}

\begin{figure} 
%
%
\begin{psfrags}%
\psfragscanon%
%
\psfrag{s05}[b][b]{\fontsize{12}{18}\fontseries{m}\mathversion{normal}\fontshape{n}\selectfont \color[rgb]{0,0,0}\setlength{\tabcolsep}{0pt}\begin{tabular}{c}$T_{{\beta}^{-}} (ms)$\end{tabular}}%
\psfrag{s06}[t][t]{\fontsize{12}{18}\fontseries{m}\mathversion{normal}\fontshape{n}\selectfont \color[rgb]{0,0,0}\setlength{\tabcolsep}{0pt}\begin{tabular}{c}MASS NUMBER\end{tabular}}%
\psfrag{s09}[b][b]{\fontsize{16}{24}\fontseries{m}\mathversion{normal}\fontshape{n}\selectfont \color[rgb]{0,0,0}\setlength{\tabcolsep}{0pt}\begin{tabular}{c}N = 82\end{tabular}}%
\psfrag{s10}[][]{\fontsize{10}{15}\fontseries{m}\mathversion{normal}\fontshape{n}\selectfont \color[rgb]{0,0,0}\setlength{\tabcolsep}{0pt}\begin{tabular}{c} \end{tabular}}%
\psfrag{s11}[][]{\fontsize{10}{15}\fontseries{m}\mathversion{normal}\fontshape{n}\selectfont \color[rgb]{0,0,0}\setlength{\tabcolsep}{0pt}\begin{tabular}{c} \end{tabular}}%
\psfrag{s12}[l][l]{\fontsize{11}{16}\fontseries{m}\mathversion{normal}\fontshape{n}\selectfont \color[rgb]{0,0,0}SM~\cite{Garcia}}%
\psfrag{s13}[l][l]{\fontsize{11}{16}\fontseries{m}\mathversion{normal}\fontshape{n}\selectfont \color[rgb]{0,0,0}NUBASE03~\cite{NUBASE03}}%
\psfrag{s14}[l][l]{\fontsize{11}{16}\fontseries{m}\mathversion{normal}\fontshape{n}\selectfont \color[rgb]{0,0,0}ANN~\cite{Costiris09}}%
\psfrag{s15}[l][l]{\fontsize{11}{16}\fontseries{m}\mathversion{normal}\fontshape{n}\selectfont \color[rgb]{0,0,0}$pn$QRPA+\textit{ff}GT~\cite{Moller}}%
\psfrag{s16}[l][l]{\fontsize{11}{16}\fontseries{m}\mathversion{normal}\fontshape{n}\selectfont \color[rgb]{0,0,0}DF3+CQRPA~\cite{Borzov, Borzov2}}%
\psfrag{s17}[l][l]{\fontsize{11}{16}\fontseries{m}\mathversion{normal}\fontshape{n}\selectfont \color[rgb]{0,0,0}$pn$QRPA+\textit{ff}~\cite{Marketin}}%
\psfrag{s18}[l][l]{\fontsize{11}{16}\fontseries{m}\mathversion{normal}\fontshape{n}\selectfont \color[rgb]{0,0,0}SM~\cite{Garcia}}%
%
\fontsize{12}{18}\fontseries{m}\mathversion{normal}%
\fontshape{n}\selectfont%
%
\psfrag{x01}[t][t]{$120$}%
\psfrag{x02}[t][t]{$125$}%
\psfrag{x03}[t][t]{$130$}%
\psfrag{x04}[t][t]{$135$}%
%
\psfrag{v01}[r][r]{$10^{0}$}%
\psfrag{v02}[r][r]{$10^{2}$}%
\psfrag{v03}[r][r]{$10^{4}$}%
\psfrag{v04}[r][r]{$10^{6}$}%
\psfrag{v05}[r][r]{$10^{8}$}%
%
\resizebox{9cm}{!}{\includegraphics{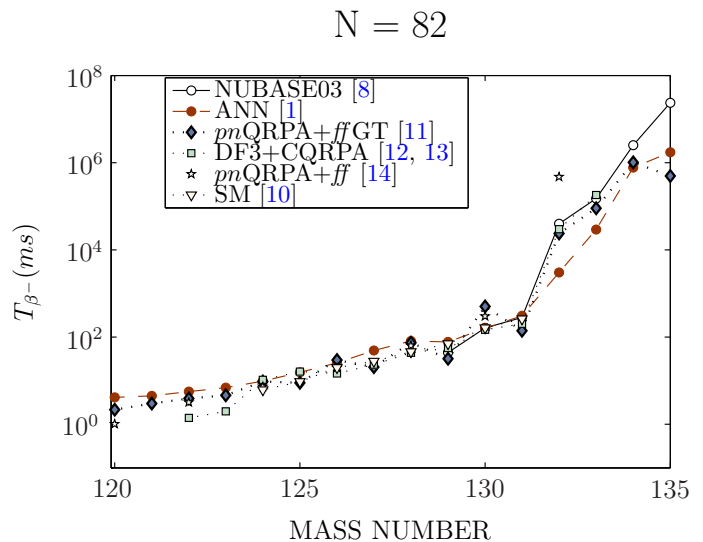}}%
\end{psfrags}%
%

\caption{\label{fig:n822}(Color online) Beta-decay half-lives 
$T_{\beta^{-}}$ produced by the standard ANN model for the r-ladder 
isotonic chain at $N=82$ (from $Z=38$ up to $Z=53$) in comparison 
with experimental values~\cite{NUBASE03}, and with corresponding 
results from $pn$QRPA+\textit{ff}GT~\cite{Moller}, 
DF3+CQRPA~\cite{Borzov, Borzov2}, $pn$-RQRPA+\textit{ff}~\cite{Marketin} 
(only for even-even nuclei), and SM~\cite{Garcia} calculations.}
\end{figure}

Reproduction of the abundances at the second peak (around $A \simeq 130$) 
is a major goal of theoretical research on r-process nucleosynthesis 
directly related to studies of $\beta$-decay of nuclei in the region 
of the $N=82$ closed neutron shell.  Each $N=82$ r-process isotope acts 
as a waiting-point for different ranges of neutron density.  Such 
studies have reached the r-process elements ($\rm{Ag}$, $\rm{Cd}$) 
and right wing ($\rm{In}$, $\rm{Sn}$, $\rm{Sb}$, $\rm{Te}$, $\rm{I}$) 
at the top of the $A \simeq 130$ r-process peak. Recently $\beta^-$ 
half-lives of very neutron-rich nuclides in the left wing have 
been determined experimentally, but measurements have not yet reached 
the r-process path in this region.  The left-wing region of nuclei,
with $A\simeq 110 - 125$ up to the onset of the $A\simeq 130$ peak,
is an especially interesting one for r-process physics, in the sense
that within this range most astrophysical r-process models overestimate 
solar system isotopic abundances by an order of magnitude or more.
This unrealistic trough has been attributed to a possible quenching 
of the $N=82$ shell gap far from stability and to possible neutrino 
spallation effects~\cite{Pfeiffer}. In addition, there are indications 
from the abundance pattern in r-process-enhanced extremely metal-poor 
stars~\cite{Terasawa, Sneden} that a second r-process is needed 
to explain the solar abundance pattern in the $A = 90 - 130$ mass region. 
Moreover, galactic chemical models of the slow neutron capture process in 
stars on the asymptotic giant branch imply that at least one additional 
primary s-process is required in this mass region~\cite{Travaglio}. 
Good nuclear inputs, including  $\beta^-$ half-lives, are 
therefore required in order to distinguish the different neutron-capture 
processes and extract reliable information on them.  
 
\begin{figure} [bth]
%
%
\begin{psfrags}%
\psfragscanon%
%
\psfrag{s05}[b][b]{\fontsize{12}{18}\fontseries{m}\mathversion{normal}\fontshape{n}\selectfont \color[rgb]{0,0,0}\setlength{\tabcolsep}{0pt}\begin{tabular}{c}$T_{{\beta}^{-}} (ms)$\end{tabular}}%
\psfrag{s06}[t][t]{\fontsize{12}{18}\fontseries{m}\mathversion{normal}\fontshape{n}\selectfont \color[rgb]{0,0,0}\setlength{\tabcolsep}{0pt}\begin{tabular}{c}MASS NUMBER\end{tabular}}%
\psfrag{s07}[b][b]{\fontsize{16}{24}\fontseries{m}\mathversion{normal}\fontshape{n}\selectfont \color[rgb]{0,0,0}\setlength{\tabcolsep}{0pt}\begin{tabular}{c}$_{45}\rm{Rh}$\end{tabular}}%
\psfrag{s10}[][]{\fontsize{10}{15}\fontseries{m}\mathversion{normal}\fontshape{n}\selectfont \color[rgb]{0,0,0}\setlength{\tabcolsep}{0pt}\begin{tabular}{c} \end{tabular}}%
\psfrag{s11}[][]{\fontsize{10}{15}\fontseries{m}\mathversion{normal}\fontshape{n}\selectfont \color[rgb]{0,0,0}\setlength{\tabcolsep}{0pt}\begin{tabular}{c} \end{tabular}}%
\psfrag{s12}[l][l]{\fontsize{11}{16}\fontseries{m}\mathversion{normal}\fontshape{n}\selectfont \color[rgb]{0,0,0}DF3+CQRPA~\cite{Borzov, Borzov2}}%
\psfrag{s13}[l][l]{\fontsize{11}{16}\fontseries{m}\mathversion{normal}\fontshape{n}\selectfont \color[rgb]{0,0,0}NUBASE03~\cite{NUBASE03}}%
\psfrag{s14}[l][l]{\fontsize{11}{16}\fontseries{m}\mathversion{normal}\fontshape{n}\selectfont \color[rgb]{0,0,0}New Exp.~\cite{Montes2}}%
\psfrag{s15}[l][l]{\fontsize{11}{16}\fontseries{m}\mathversion{normal}\fontshape{n}\selectfont \color[rgb]{0,0,0}ANN~\cite{Costiris09}}%
\psfrag{s16}[l][l]{\fontsize{11}{16}\fontseries{m}\mathversion{normal}\fontshape{n}\selectfont \color[rgb]{0,0,0}$pn$QRPA+\textit{ff}GT~\cite{Moller}}%
\psfrag{s17}[l][l]{\fontsize{11}{16}\fontseries{m}\mathversion{normal}\fontshape{n}\selectfont \color[rgb]{0,0,0}DF3+CQRPA~\cite{Borzov, Borzov2}}%
%
\fontsize{12}{18}\fontseries{m}\mathversion{normal}%
\fontshape{n}\selectfont%
%
\psfrag{x01}[t][t]{$105$}%
\psfrag{x02}[t][t]{$110$}%
\psfrag{x03}[t][t]{$115$}%
\psfrag{x04}[t][t]{$120$}%
\psfrag{x05}[t][t]{$125$}%
%
\psfrag{v01}[r][r]{$10^{2}$}%
\psfrag{v02}[r][r]{$10^{4}$}%
\psfrag{v03}[r][r]{$10^{6}$}%
\psfrag{v04}[r][r]{$10^{8}$}%
\psfrag{v05}[r][r]{$10^{10}$}%
%
\resizebox{9cm}{!}{\includegraphics{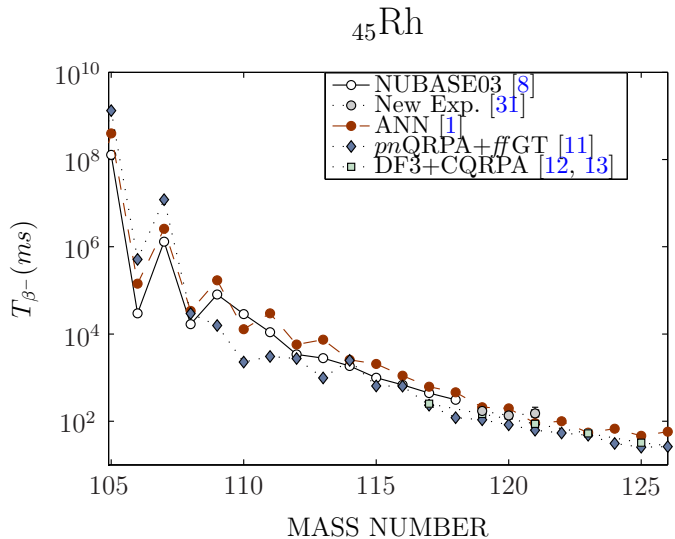}}%
\end{psfrags}%
%

\caption{\label{fig:rh45}(Color online) Beta-decay half-lives 
$T_{\beta^{-}}$ produced by the ANN model for the isotopic chain of 
$_{45}{\rm Rh}$ (from $N=60$ up to $N=81$) in comparison with experimental 
values~\cite{NUBASE03, Montes2}, and with corresponding results from  
$pn$QRPA+\textit{ff}GT~\cite{Moller} and DF3+CQRPA~\cite{Borzov, Borzov2} 
calculations.}
\end{figure}

\begin{figure} [bth]
%
%
\begin{psfrags}%
\psfragscanon%
%
\psfrag{s05}[b][b]{\fontsize{12}{18}\fontseries{m}\mathversion{normal}\fontshape{n}\selectfont \color[rgb]{0,0,0}\setlength{\tabcolsep}{0pt}\begin{tabular}{c}$T_{{\beta}^{-}} (ms)$\end{tabular}}%
\psfrag{s06}[t][t]{\fontsize{12}{18}\fontseries{m}\mathversion{normal}\fontshape{n}\selectfont \color[rgb]{0,0,0}\setlength{\tabcolsep}{0pt}\begin{tabular}{c}MASS NUMBER\end{tabular}}%
\psfrag{s07}[b][b]{\fontsize{16}{24}\fontseries{m}\mathversion{normal}\fontshape{n}\selectfont \color[rgb]{0,0,0}\setlength{\tabcolsep}{0pt}\begin{tabular}{c}$_{46}\rm{Pd}$\end{tabular}}%
\psfrag{s10}[][]{\fontsize{10}{15}\fontseries{m}\mathversion{normal}\fontshape{n}\selectfont \color[rgb]{0,0,0}\setlength{\tabcolsep}{0pt}\begin{tabular}{c} \end{tabular}}%
\psfrag{s11}[][]{\fontsize{10}{15}\fontseries{m}\mathversion{normal}\fontshape{n}\selectfont \color[rgb]{0,0,0}\setlength{\tabcolsep}{0pt}\begin{tabular}{c} \end{tabular}}%
\psfrag{s12}[l][l]{\fontsize{11}{16}\fontseries{m}\mathversion{normal}\fontshape{n}\selectfont \color[rgb]{0,0,0}$pn$RQRPA+\textit{ff}~\cite{Marketin}}%
\psfrag{s13}[l][l]{\fontsize{11}{16}\fontseries{m}\mathversion{normal}\fontshape{n}\selectfont \color[rgb]{0,0,0}NUBASE03~\cite{NUBASE03}}%
\psfrag{s14}[l][l]{\fontsize{11}{16}\fontseries{m}\mathversion{normal}\fontshape{n}\selectfont \color[rgb]{0,0,0}New Exp.~\cite{Montes2}}%
\psfrag{s15}[l][l]{\fontsize{11}{16}\fontseries{m}\mathversion{normal}\fontshape{n}\selectfont \color[rgb]{0,0,0}ANN~\cite{Costiris09}}%
\psfrag{s16}[l][l]{\fontsize{11}{16}\fontseries{m}\mathversion{normal}\fontshape{n}\selectfont \color[rgb]{0,0,0}$pn$QRPA+\textit{ff}GT~\cite{Moller}}%
\psfrag{s17}[l][l]{\fontsize{11}{16}\fontseries{m}\mathversion{normal}\fontshape{n}\selectfont \color[rgb]{0,0,0}DF3+CQRPA~\cite{Borzov, Borzov2}}%
\psfrag{s18}[l][l]{\fontsize{11}{16}\fontseries{m}\mathversion{normal}\fontshape{n}\selectfont \color[rgb]{0,0,0}$pn$RQRPA+\textit{ff}~\cite{Marketin}}%
%
\fontsize{12}{18}\fontseries{m}\mathversion{normal}%
\fontshape{n}\selectfont%
%
\psfrag{x01}[t][t]{$110$}%
\psfrag{x02}[t][t]{$115$}%
\psfrag{x03}[t][t]{$120$}%
\psfrag{x04}[t][t]{$125$}%
%
\psfrag{v01}[r][r]{$10^{2}$}%
\psfrag{v02}[r][r]{$10^{4}$}%
\psfrag{v03}[r][r]{$10^{6}$}%
\psfrag{v04}[r][r]{$10^{8}$}%
%
\resizebox{9cm}{!}{\includegraphics{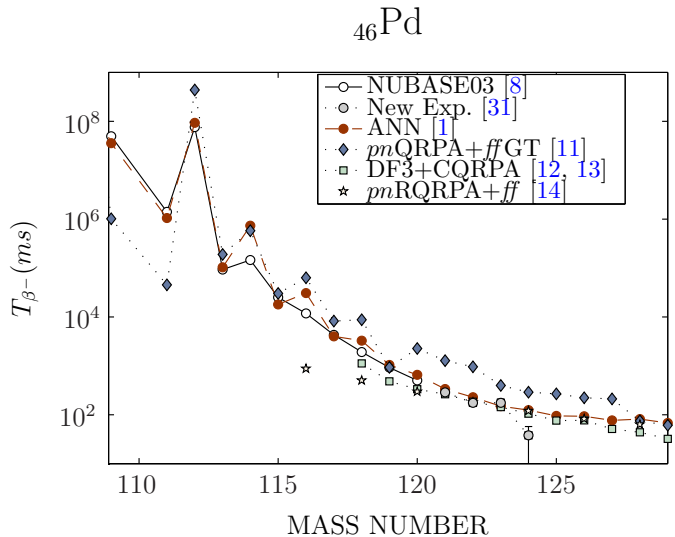}}%
\end{psfrags}%
%

\caption{\label{fig:pd46}(Color online) Beta-decay half-lives 
$T_{\beta^{-}}$ produced by the ANN model for the isotopic chain of 
$_{46}{\rm Pd}$ (from $N=63$ up to $N=83$) in comparison with experimental 
values~\cite{NUBASE03, Montes2}, and with results from   
$pn$QRPA+\textit{ff}GT~\cite{Moller}, DF3+CQRPA~\cite{Borzov, Borzov2}, 
and (only for even-even nuclei) $pn$-RQRPA+\textit{ff}~\cite{Marketin} 
calculations.}
\end{figure}

In recent years several experimental programs have focused on 
the region of the r-process path near $N=82$. 
Beta-decay properties including $\beta^-$ half-lives of exotic 
neutron-rich $\rm{Kr}$, $\rm{Sr}$, $\rm{Y}$,  $\rm{Zr}$, $\rm{Nb}$, 
$\rm{Mo}$, and $\rm{Tc}$ nuclides have been studied at the RIBF facility 
at RIKEN~\cite{Nishimura}. Additionally, the half-lives of neutron-rich 
$\rm{Ru}$, $\rm{Rh}$, and $\rm{Pd}$ nuclides have been measured at 
NSCL~\cite{Montes2}.  It also should be mentioned that half-lives for 
$N \simeq 66$ mid-shell nuclei below $A\simeq 110$, i.e., for
$^{105}\rm{Y}$, $^{106, 107}\rm{Zr}$, and $^{115}\rm{Tc}$, 
have been measured at NSCL~\cite{Pereira2009} after publication of 
NUBASE03, but we consider here the most recent RIKEN measurements.
In the chart of Fig.~\ref{fig:n82chart} the nuclides named above
are denoted by dark gray (blue) boxes, while other nuclides in this
region whose $T_{\beta^{-}}$ values have been measured earlier 
(and appear in NUBASE03~\cite{NUBASE03}) are symbolized by gray
(green) boxes.  In Table~\ref{TabIeN126} we also list $T_{\beta^{-}}$ 
values of r-process nuclides on the right wing of the $A \simeq 130$ 
peak not included in NUBASE03 but measured at CERN/ISOLDE. More 
specifically, the isotopes $^{133}\rm{Cd}$~\cite{Arndt}, 
$^{138}\rm{Sn}$~\cite{Kratz}, and $^{138-139}\rm{Sb}$~\cite{Arndt2}  
have been studied. Very neutron-rich $\rm{Cd}$, $\rm{Sn}$, and $\rm{Sb}$ 
nuclides play a critical role in r-process nucleosynthesis calculations, 
as they lie directly on the path of the r-process under a wide range 
of astrophysical conditions.  Table~\ref{TabIeN126} provides a detailed 
comparison of the outputs of the standard ANN model and theoretical 
results (as available) from the $pn$QRPA+\textit{ff}GT~\cite{Moller} 
and DF3+CQRPA~\cite{Borzov, Borzov2} calculations with the experimental 
$\beta^-$ half-lives for all the recently measured nuclides identified 
in this paragraph.  It is seen that most of the ANN predictions are 
closer to the measured values than those of the 
$pn$QRPA+\textit{ff}GT~\cite{Moller} model. This theory-thick model 
significantly overestimates the $T_{\beta^{-}}$ values for the 
relevant $\rm{Zr}$, $\rm{Nb}$, and $\rm{Mo}$ isotopes, as is 
also the case for the relevant isotopes of $\rm{Ru}$, $\rm{Pd}$, 
$\rm{Cd}$, $\rm{Sn}$, and $\rm{Sb}$.  On the other hand, the ANN 
statistical model underestimates the $\beta^-$ half-lives in almost
all cases except $\rm{Pd}$, but to a lesser extent. 

We have considered above the half-lives of neutron-rich nuclides around 
the $N=82$ r-process path that have recently been measured and therefore 
played no role in construction of the ANN model. The $T_{\beta^{-}}$ 
values of the $N=82$ isotones in the region $Z = 38 - 53$ that are 
experimentally known are displayed in Fig.~\ref{fig:n822}. In the same 
figure the values generated by the ANN model for isotopes in 
the region $Z = 38 - 53$ are compared with the corresponding 
experimental results as well as with results from 
the $pn$QRPA+\textit{ff}GT~\cite{Moller}, DF3+CQRPA~\cite{Borzov, Borzov2}, 
$pn$-RQRPA+\textit{ff}~\cite{Marketin} (only for even-even nuclei), and 
SM~\cite{Garcia} calculations.  For $Z < 50$ the $T_{\beta^{-}}$ 
values from the 
various theory-thick calculations do not differ much. It is instructive 
to further evaluate the performance of the ANN model in this regime 
by considering the half-lives of nuclides in the isotopic chain 
of $\rm{Rh}$ and $\rm{Pd}$ (Figs.~\ref{fig:rh45},~\ref{fig:pd46}). 
The $T_{\beta^{-}}$ values produced by the ANN model are generally 
in better agreement with the available experimental measurements than 
those of Ref.~\cite{Moller} and are rather closer to the corresponding 
DF3+CQRPA results~\cite{Borzov, Borzov2}.

\subsection{\label{sec:level-3-3}Nuclei in the region of the $N=126$ closed 
neutron shell (around the A$\simeq$195 peak)}

\begin{figure} 
%
%
\begin{psfrags}%
\psfragscanon%
%
\psfrag{s05}[b][b]{\fontsize{12}{18}\fontseries{m}\mathversion{normal}\fontshape{n}\selectfont \color[rgb]{0,0,0}\setlength{\tabcolsep}{0pt}\begin{tabular}{c}PROTONS (Z)\end{tabular}}%
\psfrag{s06}[t][t]{\fontsize{12}{18}\fontseries{m}\mathversion{normal}\fontshape{n}\selectfont \color[rgb]{0,0,0}\setlength{\tabcolsep}{0pt}\begin{tabular}{c}NEUTRONS (N)\end{tabular}}%
\psfrag{s09}[b][b]{\fontsize{16}{24}\fontseries{m}\mathversion{normal}\fontshape{n}\selectfont \color[rgb]{0,0,0}\setlength{\tabcolsep}{0pt}\begin{tabular}{c}Chart of nuclides up to N = 126\end{tabular}}%
\psfrag{s10}[][]{\fontsize{10}{15}\fontseries{m}\mathversion{normal}\fontshape{n}\selectfont \color[rgb]{0,0,0}\setlength{\tabcolsep}{0pt}\begin{tabular}{c} \end{tabular}}%
\psfrag{s11}[][]{\fontsize{10}{15}\fontseries{m}\mathversion{normal}\fontshape{n}\selectfont \color[rgb]{0,0,0}\setlength{\tabcolsep}{0pt}\begin{tabular}{c} \end{tabular}}%
\psfrag{s12}[l][l]{\fontsize{11}{16}\fontseries{m}\mathversion{normal}\fontshape{n}\selectfont \color[rgb]{0,0,0}Unknown}%
\psfrag{s13}[l][l]{\fontsize{11}{16}\fontseries{m}\mathversion{normal}\fontshape{n}\selectfont \color[rgb]{0,0,0}NUBASE03~\cite{NUBASE03}}%
\psfrag{s14}[l][l]{\fontsize{11}{16}\fontseries{m}\mathversion{normal}\fontshape{n}\selectfont \color[rgb]{0,0,0}New Exp.~\cite{Benlliure2}}%
\psfrag{s15}[l][l]{\fontsize{11}{16}\fontseries{m}\mathversion{normal}\fontshape{n}\selectfont \color[rgb]{0,0,0}$\beta^{+}$}%
\psfrag{s16}[l][l]{\fontsize{11}{16}\fontseries{m}\mathversion{normal}\fontshape{n}\selectfont \color[rgb]{0,0,0}Stable}%
\psfrag{s17}[l][l]{\fontsize{11}{16}\fontseries{m}\mathversion{normal}\fontshape{n}\selectfont \color[rgb]{0,0,0}Unknown}%
%
\fontsize{14}{21}\fontseries{m}\mathversion{normal}%
\fontshape{n}\selectfont%
%
\psfrag{x01}[t][t]{$112$}%
\psfrag{x02}[t][t]{$114$}%
\psfrag{x03}[t][t]{$116$}%
\psfrag{x04}[t][t]{$118$}%
\psfrag{x05}[t][t]{$120$}%
\psfrag{x06}[t][t]{$122$}%
\psfrag{x07}[t][t]{$124$}%
\psfrag{x08}[t][t]{$126$}%
%
\psfrag{v01}[r][r]{}%
\psfrag{v02}[r][r]{$_{72}\rm{Hf}$}%
\psfrag{v03}[r][r]{$_{73}\rm{Ta}$}%
\psfrag{v04}[r][r]{$_{74}\rm{W}$}%
\psfrag{v05}[r][r]{$_{75}\rm{Re}$}%
\psfrag{v06}[r][r]{$_{76}\rm{Os}$}%
\psfrag{v07}[r][r]{$_{77}\rm{Ir}$}%
\psfrag{v08}[r][r]{$_{78}\rm{Pt}$}%
\psfrag{v09}[r][r]{$_{79}\rm{Au}$}%
\psfrag{v10}[r][r]{$_{80}\rm{Hg}$}%
\psfrag{v11}[r][r]{$_{81}\rm{Tl}$}%
\psfrag{v12}[r][r]{}%
%
\resizebox{9cm}{!}{\includegraphics{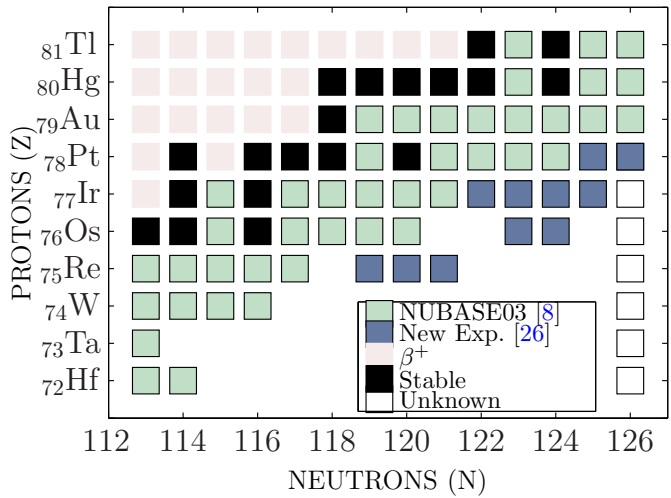}}%
\end{psfrags}%
%
  
\caption{\label{fig:n126chart}(Color online) Cluster plot of the chart 
of nuclides up to the $N=126$ closed shell. Black boxes indicate stable 
nuclides. Gray (green) boxes indicate nuclides with measured 
$\beta^{-}$-decay half-lives $T_{\beta^{-}}$ reported in 
NUBASE03~\cite{NUBASE03}, while dark gray boxes (blue) identify those 
with newly measured $T_{\beta^{-}}$ beyond NUBASE03~\cite{Benlliure2} . 
Empty boxes denote nuclides at the $N=126$ closed shell for 
which no experimental $T_{\beta^{-}}$ value is available. Theoretical 
values for $T_{\beta^{-}}$ from the standard ANN model as well 
as from conventional nuclear models are plotted in Fig.~\ref{fig:n1262}.}
\end{figure}

\begin{figure} 
%
%
\begin{psfrags}%
\psfragscanon%
%
\psfrag{s05}[b][b]{\fontsize{12}{18}\fontseries{m}\mathversion{normal}\fontshape{n}\selectfont \color[rgb]{0,0,0}\setlength{\tabcolsep}{0pt}\begin{tabular}{c}$T_{{\beta}^{-}} (ms)$\end{tabular}}%
\psfrag{s06}[t][t]{\fontsize{12}{18}\fontseries{m}\mathversion{normal}\fontshape{n}\selectfont \color[rgb]{0,0,0}\setlength{\tabcolsep}{0pt}\begin{tabular}{c}MASS NUMBER\end{tabular}}%
\psfrag{s07}[b][b]{\fontsize{16}{24}\fontseries{m}\mathversion{normal}\fontshape{n}\selectfont \color[rgb]{0,0,0}\setlength{\tabcolsep}{0pt}\begin{tabular}{c}N = 126\end{tabular}}%
\psfrag{s10}[][]{\fontsize{10}{15}\fontseries{m}\mathversion{normal}\fontshape{n}\selectfont \color[rgb]{0,0,0}\setlength{\tabcolsep}{0pt}\begin{tabular}{c} \end{tabular}}%
\psfrag{s11}[][]{\fontsize{10}{15}\fontseries{m}\mathversion{normal}\fontshape{n}\selectfont \color[rgb]{0,0,0}\setlength{\tabcolsep}{0pt}\begin{tabular}{c} \end{tabular}}%
\psfrag{s12}[l][l]{\fontsize{11}{16}\fontseries{m}\mathversion{normal}\fontshape{n}\selectfont \color[rgb]{0,0,0}DF3+CQRPA~\cite{Borzov}}%
\psfrag{s13}[l][l]{\fontsize{11}{16}\fontseries{m}\mathversion{normal}\fontshape{n}\selectfont \color[rgb]{0,0,0}NUBASE03~\cite{NUBASE03}}%
\psfrag{s14}[l][l]{\fontsize{11}{16}\fontseries{m}\mathversion{normal}\fontshape{n}\selectfont \color[rgb]{0,0,0}New Exp.~\cite{Benlliure2}}%
\psfrag{s15}[l][l]{\fontsize{11}{16}\fontseries{m}\mathversion{normal}\fontshape{n}\selectfont \color[rgb]{0,0,0}ANN~\cite{Costiris09}}%
\psfrag{s16}[l][l]{\fontsize{11}{16}\fontseries{m}\mathversion{normal}\fontshape{n}\selectfont \color[rgb]{0,0,0}$pn$QRPA+\textit{ff}GT~\cite{Moller}}%
\psfrag{s17}[l][l]{\fontsize{11}{16}\fontseries{m}\mathversion{normal}\fontshape{n}\selectfont \color[rgb]{0,0,0}DF3+CQRPA~\cite{Borzov}}%
%
\fontsize{12}{18}\fontseries{m}\mathversion{normal}%
\fontshape{n}\selectfont%
%
\psfrag{x01}[t][t]{$198$}%
\psfrag{x02}[t][t]{$200$}%
\psfrag{x03}[t][t]{$202$}%
\psfrag{x04}[t][t]{$204$}%
\psfrag{x05}[t][t]{$206$}%
%
\psfrag{v01}[r][r]{$10^{2}$}%
\psfrag{v02}[r][r]{$10^{3}$}%
\psfrag{v03}[r][r]{$10^{4}$}%
\psfrag{v04}[r][r]{$10^{5}$}%
\psfrag{v05}[r][r]{$10^{6}$}%
\psfrag{v06}[r][r]{$10^{7}$}%
\psfrag{v07}[r][r]{$10^{8}$}%
%
\resizebox{9cm}{!}{\includegraphics{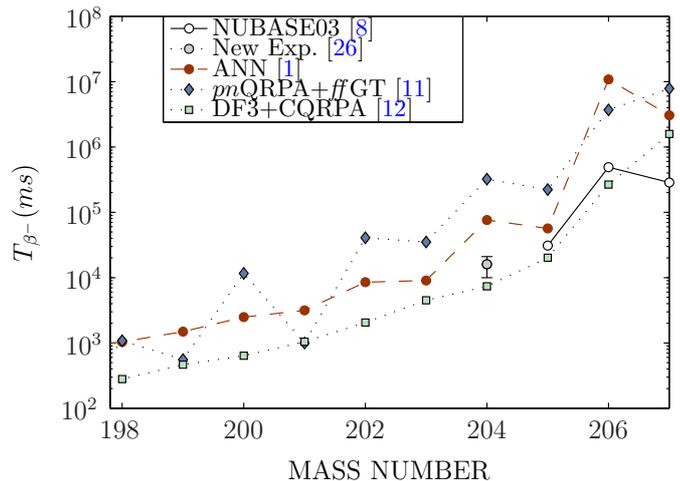}}%
\end{psfrags}%
%

\caption{\label{fig:n1262}(Color online) Beta-decay half-lives 
$T_{\beta^{-}}$ produced by the ANN model for the r-ladder isotonic chain 
at $N=126$ (from $Z=72$ up to $Z=81$) in comparison with experimental 
values~\cite{NUBASE03, Benlliure2}, and with results from 
$pn$QRPA+\textit{ff}GT~\cite{Moller} and DF3+CQRPA~\cite{Borzov} 
calculations.}
\end{figure}

The half-lives of heavy neutron-rich nuclides in the $N=126$ region 
considerably below the doubly magic nucleus $^{208}\rm{Pb}$ are also 
important for an understanding of the r-process. They play a key
role in determining the relative $A= 195$ r-process peak abundances. 
They also determine how rapidly the heaviest nuclei are synthesized during 
the r-process (and hence the strength of ``fission cycling''~\cite{Nieto}), 
as well as the r-process end point.

\begin{figure} [bth]
%
%
\begin{psfrags}%
\psfragscanon%
%
\psfrag{s05}[b][b]{\fontsize{12}{18}\fontseries{m}\mathversion{normal}\fontshape{n}\selectfont \color[rgb]{0,0,0}\setlength{\tabcolsep}{0pt}\begin{tabular}{c}$T_{{\beta}^{-}} (ms)$\end{tabular}}%
\psfrag{s06}[t][t]{\fontsize{12}{18}\fontseries{m}\mathversion{normal}\fontshape{n}\selectfont \color[rgb]{0,0,0}\setlength{\tabcolsep}{0pt}\begin{tabular}{c}MASS NUMBER\end{tabular}}%
\psfrag{s07}[b][b]{\fontsize{16}{24}\fontseries{m}\mathversion{normal}\fontshape{n}\selectfont \color[rgb]{0,0,0}\setlength{\tabcolsep}{0pt}\begin{tabular}{c}$_{76}\rm{Os}$\end{tabular}}%
\psfrag{s10}[][]{\fontsize{10}{15}\fontseries{m}\mathversion{normal}\fontshape{n}\selectfont \color[rgb]{0,0,0}\setlength{\tabcolsep}{0pt}\begin{tabular}{c} \end{tabular}}%
\psfrag{s11}[][]{\fontsize{10}{15}\fontseries{m}\mathversion{normal}\fontshape{n}\selectfont \color[rgb]{0,0,0}\setlength{\tabcolsep}{0pt}\begin{tabular}{c} \end{tabular}}%
\psfrag{s12}[l][l]{\fontsize{11}{16}\fontseries{m}\mathversion{normal}\fontshape{n}\selectfont \color[rgb]{0,0,0}DF3+CQRPA~\cite{Borzov}}%
\psfrag{s13}[l][l]{\fontsize{11}{16}\fontseries{m}\mathversion{normal}\fontshape{n}\selectfont \color[rgb]{0,0,0}NUBASE03~\cite{NUBASE03}}%
\psfrag{s14}[l][l]{\fontsize{11}{16}\fontseries{m}\mathversion{normal}\fontshape{n}\selectfont \color[rgb]{0,0,0}New Exp.~\cite{Benlliure2}}%
\psfrag{s15}[l][l]{\fontsize{11}{16}\fontseries{m}\mathversion{normal}\fontshape{n}\selectfont \color[rgb]{0,0,0}ANN~\cite{Costiris09}}%
\psfrag{s16}[l][l]{\fontsize{11}{16}\fontseries{m}\mathversion{normal}\fontshape{n}\selectfont \color[rgb]{0,0,0}$pn$QRPA+\textit{ff}GT~\cite{Moller}}%
\psfrag{s17}[l][l]{\fontsize{11}{16}\fontseries{m}\mathversion{normal}\fontshape{n}\selectfont \color[rgb]{0,0,0}DF3+CQRPA~\cite{Borzov}}%
%
\fontsize{12}{18}\fontseries{m}\mathversion{normal}%
\fontshape{n}\selectfont%
%
\psfrag{x01}[t][t]{$196$}%
\psfrag{x02}[t][t]{$198$}%
\psfrag{x03}[t][t]{$200$}%
\psfrag{x04}[t][t]{$202$}%
\psfrag{x05}[t][t]{$204$}%
\psfrag{x06}[t][t]{$206$}%
%
\psfrag{v01}[r][r]{$10^{2}$}%
\psfrag{v02}[r][r]{$10^{3}$}%
\psfrag{v03}[r][r]{$10^{4}$}%
\psfrag{v04}[r][r]{$10^{5}$}%
\psfrag{v05}[r][r]{$10^{6}$}%
\psfrag{v06}[r][r]{$10^{7}$}%
\psfrag{v07}[r][r]{$10^{8}$}%
%
\resizebox{9cm}{!}{\includegraphics{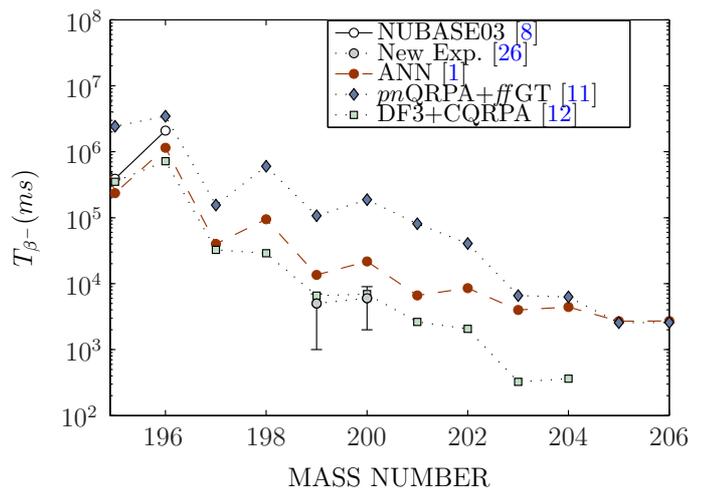}}%
\end{psfrags}%
%

\caption{\label{fig:os76}(Color online) Beta-decay half-lives 
$T_{\beta^{-}}$ produced by the ANN model for the isotopic chain of 
$\rm{_{76}Os}$ (from $N=119$ up to $N=130$) in comparison with experimental 
values~\cite{NUBASE03, Benlliure2}, and with results from 
$pn$QRPA+\textit{ff}GT~\cite{Moller} and DF3+CQRPA~\cite{Borzov} calculations.}
\end{figure}

\begin{figure} [bth]
%
%
\begin{psfrags}%
\psfragscanon%
%
\psfrag{s05}[b][b]{\fontsize{12}{18}\fontseries{m}\mathversion{normal}\fontshape{n}\selectfont \color[rgb]{0,0,0}\setlength{\tabcolsep}{0pt}\begin{tabular}{c}$T_{{\beta}^{-}} (ms)$\end{tabular}}%
\psfrag{s06}[t][t]{\fontsize{12}{18}\fontseries{m}\mathversion{normal}\fontshape{n}\selectfont \color[rgb]{0,0,0}\setlength{\tabcolsep}{0pt}\begin{tabular}{c}MASS NUMBER\end{tabular}}%
\psfrag{s07}[b][b]{\fontsize{16}{24}\fontseries{m}\mathversion{normal}\fontshape{n}\selectfont \color[rgb]{0,0,0}\setlength{\tabcolsep}{0pt}\begin{tabular}{c}$_{77}\rm{Ir}$\end{tabular}}%
\psfrag{s10}[][]{\fontsize{10}{15}\fontseries{m}\mathversion{normal}\fontshape{n}\selectfont \color[rgb]{0,0,0}\setlength{\tabcolsep}{0pt}\begin{tabular}{c} \end{tabular}}%
\psfrag{s11}[][]{\fontsize{10}{15}\fontseries{m}\mathversion{normal}\fontshape{n}\selectfont \color[rgb]{0,0,0}\setlength{\tabcolsep}{0pt}\begin{tabular}{c} \end{tabular}}%
\psfrag{s12}[l][l]{\fontsize{11}{16}\fontseries{m}\mathversion{normal}\fontshape{n}\selectfont \color[rgb]{0,0,0}DF3+CQRPA~\cite{Borzov}}%
\psfrag{s13}[l][l]{\fontsize{11}{16}\fontseries{m}\mathversion{normal}\fontshape{n}\selectfont \color[rgb]{0,0,0}NUBASE03~\cite{NUBASE03}}%
\psfrag{s14}[l][l]{\fontsize{11}{16}\fontseries{m}\mathversion{normal}\fontshape{n}\selectfont \color[rgb]{0,0,0}New Exp.~\cite{Benlliure2}}%
\psfrag{s15}[l][l]{\fontsize{11}{16}\fontseries{m}\mathversion{normal}\fontshape{n}\selectfont \color[rgb]{0,0,0}ANN~\cite{Costiris09}}%
\psfrag{s16}[l][l]{\fontsize{11}{16}\fontseries{m}\mathversion{normal}\fontshape{n}\selectfont \color[rgb]{0,0,0}$pn$QRPA+\textit{ff}GT~\cite{Moller}}%
\psfrag{s17}[l][l]{\fontsize{11}{16}\fontseries{m}\mathversion{normal}\fontshape{n}\selectfont \color[rgb]{0,0,0}DF3+CQRPA~\cite{Borzov}}%
%
\fontsize{12}{18}\fontseries{m}\mathversion{normal}%
\fontshape{n}\selectfont%
%
\psfrag{x01}[t][t]{$194$}%
\psfrag{x02}[t][t]{$196$}%
\psfrag{x03}[t][t]{$198$}%
\psfrag{x04}[t][t]{$200$}%
\psfrag{x05}[t][t]{$202$}%
\psfrag{x06}[t][t]{$204$}%
\psfrag{x07}[t][t]{$206$}%
%
\psfrag{v01}[r][r]{$10^{2}$}%
\psfrag{v02}[r][r]{$10^{4}$}%
\psfrag{v03}[r][r]{$10^{6}$}%
\psfrag{v04}[r][r]{$10^{8}$}%
\psfrag{v05}[r][r]{$10^{10}$}%
%
\resizebox{9cm}{!}{\includegraphics{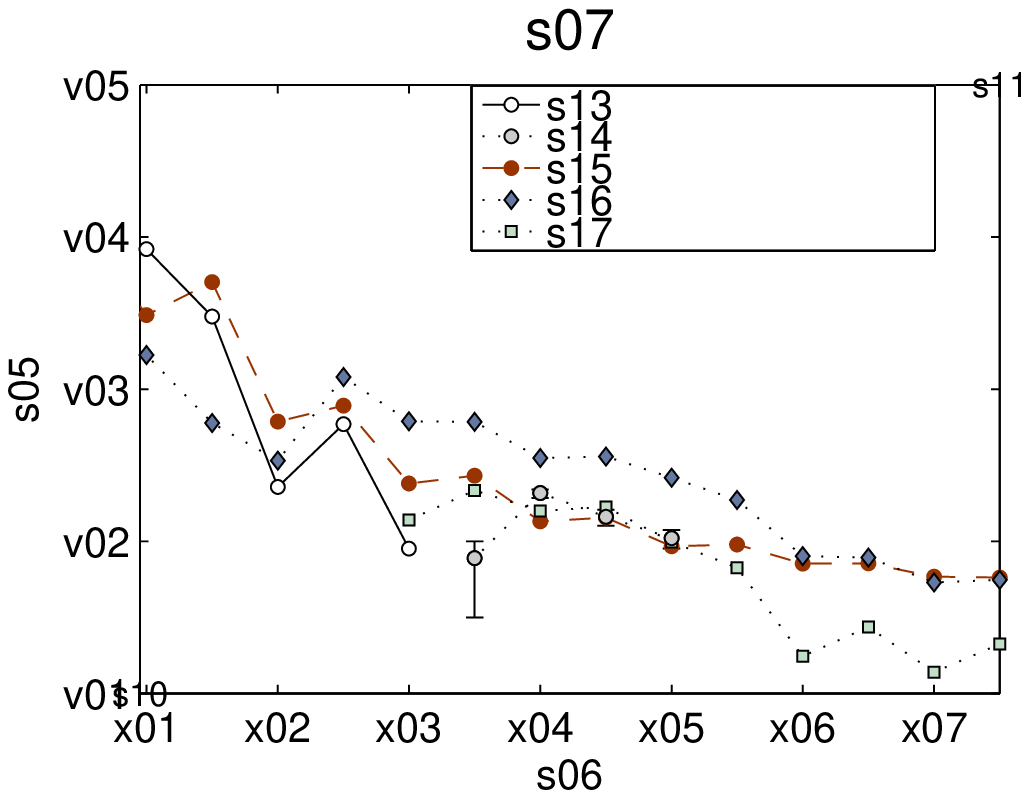}}%
\end{psfrags}%
%

\caption{\label{fig:ir77}(Color online) Beta-decay half-lives $T_{\beta^{-}}$ 
given by the ANN model for the isotopic chain of $\rm{_{77}Ir}$ (from $N=117$ 
up to $N=130$) in comparison with experimental values~\cite{NUBASE03, 
Benlliure2}, and with results from $pn$QRPA+\textit{ff}GT~\cite{Moller} 
and DF3+CQRPA~\cite{Borzov} calculations.}
\end{figure}

Recently, experimental results in the $N=126$ region obtained by cold
fragmentation and a novel method of analysis~\cite{Benlliure2, Nieto} 
have been reported by the RISING Collaboration (projects S227, S312) at 
GSI.  There is the prospect of experimental measurements for more 
neutron-rich nuclides in this region with the advent of a new 
generation of in-flight secondary beam facilities, i.e., FAIR at 
GSI, RIBF at RIKEN, and FRIB at MSU~\cite{Facilities}. 

A chart of nuclides up to the $N=126$ closed shell with known 
experimental $\beta^-$-decay half-lives is displayed in
Fig.~\ref{fig:n126chart}.  Nuclides whose $T_{\beta^{-}}$ values 
are included in NUBASE03~\cite{NUBASE03} are represented by 
gream GSI~\cite{Alvarez, Nieto, Benlliure2} after publication of 
NUBASE03 are indicated in dark gray (blue).  Specifically, these 
latter experimental half-lives include measurements for 
$^{194-196}\rm{Re}$, $^{199-200}\rm{Os}$, $^{199-202}\rm{Ir}$, 
and $^{203-204}\rm{Pt}$ and provide evidence for tests of ANN
and traditional models of beta-decay systematics.  Table~\ref{TabIeN126} 
juxtaposes the predictions from the standard ANN model and from 
$pn$QRPA+\textit{ff}GT~\cite{Moller} and DF3+CQRPA~\cite{Borzov} 
calculations, with the recent half-life data from GSI.  In 
particular, the DF3+CQRPA calculation follows that for N$\simeq$126 
described in Ref.~\cite{Borzov}, being based on 
the Fayans energy-density functional but with no energy-dependent
smearing in the treatment of Gamow-Teller and first-forbidden 
transitions. 
Fig.~\ref{fig:n1262} compares the ANN predictions for the isotonic 
chain at $N=126$ (from $Z=72$ up to 81) with available experimental 
values and with results from the $pn$QRPA+\textit{ff}GT~\cite{Moller} 
and DF3+CQRPA calculations~\cite{Borzov}.  The ANN results are seen
to be in better agreement with the experimental data than those
of M\"oller et al.~\cite{Moller} (which for most of the nuclei 
are more than an order of magnitude larger than the data), and they 
show less pronounced odd-even staggering. Generally, the ANN model 
predicts shorter half-lives for $N=126$ isotones than the 
$pn$QRPA+\textit{ff}GT model, thus implying that the matter flow 
to the heavier fissioning nuclides is faster.  Compared to the results 
from the DF3+CQRPA calculations~\cite{Borzov}, the ANN values are 
generally somewhat larger. The performance of our model in this 
region can be further evaluated by studying the isotopic chains of 
$\rm{Os}$ and $\rm{Ir}$ (see Figs.~\ref{fig:os76} and~\ref{fig:ir77}, 
respectively). Generally, half-lives predicted by the ANN model 
are quite close to the experimental values, longer than those of 
the DF3+CQRPA calculations~\cite{Borzov} but shorter than those 
of the $pn$QRPA+\textit{ff}GT~\cite{Moller} model, the latter 
behavior being exhibited especially for $\rm{Os}$ isotopes with 
$A\leq 202$ and for $\rm{Ir}$ isotopes with $197 \leq A \leq 
203$. (One could say that the ANN interpolates between the two 
theory-thick models.) The good agreement of the results of the 
ANN model with the experimental half-lives in this region of the 
nuclear chart, as well as with those in the $N=50$ and $N=82$ 
regions, gives some confidence in the values extracted for the 
$N=126$ r-process waiting-points.

In the previous sections the ANN model has been employed (i) to
generate statistically derived values of the half-lives of 
neutron-rich nuclei close to the waiting-point nuclei at $N=50$, 
82, and 126 that have recently been measured, and (i) to predict 
the half-lives of the rest of the waiting-point nuclei at $N=50$, 
82, and 126.  Experimental values of $T_{\beta^{-}}$ for nuclei 
 not included in NUBASE03, along 
with the predicted half-life values produced by the ANN model and 
calculated values from two conventional nuclear models
($pn$QRPA+\textit{ff}GT~\cite{Moller} and DF3+CQRPA~\cite{Borzov}) 
are also listed in Table~\ref{TabIeN126}.  Based on the reported 
values of the error measure $\sigma_{\rm rms}$ (0.37 for the ANN 
model compared to 0.64 for the model of Ref.~\cite{Moller}), 
rather satisfactory predictive performance has been achieved 
with the ANN model.

\begin{longtable*}{lllll}
& \multicolumn{4}{c}{  $T_{\beta^{-}} (ms)$} \\ 
\cline{2-5}
 Nucl.  & Exp. Data&  ANN Model~\cite{Costiris09}  & $pn$QRPA+\textit{ff}GT~\cite{Moller} & DF3+CQRPA~\cite{Borzov}  \\
\toprule 
 \multicolumn{5}{c}{Nuclides  around the  N=50 shell closure}\\  
\hline
$^{61}\rm{Ti}$ &  $  15\pm 4$~\cite{Daugas}&   27.65   &  22.18  &3.4\\ 
$^{64}\rm{V}$ &  $  19\pm 8$~\cite{Daugas}&    23.38 &   7.62  &3.7\\  
$^{71}\rm{Fe}$ &  $28\pm 15$~\cite{Daugas}&   66 &  148&-\\ 
$^{73}\rm{Co}$ &  $41\pm 4$~\cite{Daugas}&    104& 31 &- \\ 
$^{74}\rm{Co}$ &  $19\pm 7$~\cite{Daugas}&    89&  23&- \\ 
$^{75}\rm{Co}$ &  $30\pm 11$~\cite{Hosmer2010}&  55  & 19 & -\\ 
$^{77}\rm{Ni}$ &  $128  {^{+27}_{-33}}$~\cite{Hosmer}&  79 & 208 &144\\ 
$^{78}\rm{Ni}$ &  $110  {^{+100}_{-60}}$~\cite{Hosmer}&  57 & 224 &108\\ 
$^{80}\rm{Cu}$ &  $170 {^{+110}_{-50}}$~\cite{Hosmer2010}&116   &  85& -\\ 
\hline
 \multicolumn{5}{c}{Nuclides  around the  N=82 shell closure} \\ 
\hline
$^{100}\rm{Kr}$ &  $  7{^{+ 11}_{- 3}}$~\cite{Nishimura}&   16&  42&-\\ 
$^{103}\rm{Sr}$ &  $  68{^{+ 48}_{- 20}}$~\cite{Nishimura}&   47&  33&-\\  
$^{104}\rm{Sr}$ &  $  43{^{+9 }_{-7 }}$~\cite{Nishimura}&   23&  71&-\\ 
$^{105}\rm{Sr}$ &  $  40{^{+ 36}_{- 13}}$~\cite{Nishimura}&   21&  45&-\\ 
$^{105}\rm{Y}$  &   $83{^{+5 }_{-4 }} $~\cite{Nishimura} &       58& 46 &-\\ 
$^{106}\rm{Y}$ &  $  62{^{+25 }_{- 14}}$~\cite{Nishimura}&  61 &  34&-\\  
$^{107}\rm{Y}$ &  $  41{^{+ 15}_{- 9}}$~\cite{Nishimura}&   26&  29&-\\ 
$^{108}\rm{Y}$ &  $  25{^{+ 66}_{- 10}}$~\cite{Nishimura}&  33 &  22&-  \\ 
$^{106}\rm{Zr}$     & $186 {^{+11}_{-10}} $~\cite{Nishimura} &         106& 322&-    \\ 
$^{107}\rm{Zr}$     & $138\pm4 $~\cite{Nishimura} &       75  & 177 &-\\ 
$^{108}\rm{Zr}$ &  $  73\pm4$~\cite{Nishimura}&   36& 162 &-\\ 
$^{109}\rm{Zr}$ &  $  63{^{+38 }_{- 17}}$~\cite{Nishimura}&   32&  108&-\\ 
$^{110}\rm{Zr}$ &  $  37{^{+ 17}_{- 9}}$~\cite{Nishimura}&   16&  76&-\\ 
$^{111}\rm{Nb}$ &  $ 51 {^{+6 }_{- 5}}$~\cite{Nishimura}&   42&  151&-\\ 
$^{112}\rm{Nb}$ &  $  33{^{+9 }_{-6 }}$~\cite{Nishimura}&  51 & 75 &-   \\ 
$^{111}\rm{Mo}$     & $200  {^{+40}_{-35}} $~\cite{Pereira2009}  &  145  & 808& 146 \\  
$^{112}\rm{Mo}$ &  $  120{^{+13 }_{-11 }}$~\cite{Nishimura}&   71&  581 &- \\ 
$^{113}\rm{Mo}$ &  $  78{^{+ 6}_{- 5}}$~\cite{Nishimura}&  58 &  121&- \\ 
$^{114}\rm{Mo}$ &  $  60{^{+13 }_{-9 }}$~\cite{Nishimura}&   29&  103 &-\\ 
$^{115}\rm{Mo}$ &  $  51{^{+ 79}_{- 19}}$~\cite{Nishimura}&   27&  49   &-\\ 
$^{115}\rm{Tc}^*$ &   $83  {^{+20}_{-13}} $~\cite{Nishimura}   &       84&    71 & 134  \\  
$^{116}\rm{Tc}$ &  $  56{^{+ 15}_{- 10}}$~\cite{Nishimura}&   95& 44 &- \\ 
$^{117}\rm{Tc}$ &  $ 89 {^{+ 95}_{- 30}}$~\cite{Nishimura}&   37&  40 &-\\  
$^{116}\rm{Ru}$ &  $ 204  {^{+32}_{-29}} $~\cite{Montes2}   &     188&      540 &  193 \\ 
$^{117}\rm{Ru}$ &   $ 142 {^{+18}_{-17}} $~\cite{Montes2}  &     129&      163 &127   \\   
$^{118}\rm{Ru}^*$ &   $ 123  {^{+48}_{-35}} $~\cite{Montes2}   &    69 &    212& 95 \\ 
$^{119}\rm{Rh}$ &   $ 171\pm 18 $~\cite{Montes2} &     209&    108  & 146 \\ 
$^{120}\rm{Rh}^*$ &   $136  {^{+14}_{-13}}$~\cite{Montes2} &     196&   83 & -   \\ 
$^{121}\rm{Rh}^*$ &   $ 151  {^{+67}_{-58}} $~\cite{Montes2}&    91 &    62 &  87  \\ 
$^{121}\rm{Pd}$ &   $ 285\pm 24 $~\cite{Montes2} &334&    1275& 262\\  
$^{122}\rm{Pd}^*$&   $ 175\pm 16$~\cite{Montes2}  &  227&    951   &  184\\ 
$^{123}\rm{Pd}$ &   $174  {^{+38}_{-34}} $~\cite{Montes2}    & 149   &  397  &   143 \\  
$^{124}\rm{Pd}^*$ &   $38  {^{+38}_{-19}} $~\cite{Montes2} &  124   &  289   & 105   \\ 
$^{133}\rm{Cd}$ &   $ 57 \pm 10 $~\cite{Arndt}&    57 &   185  & 47    \\  
$^{138}\rm{Sn}$ &   $ 150 \pm 60 $~\cite{Kratz}&    113 &   336  & 240    \\
$^{138}\rm{Sb}$ &    $350 \pm 15 $~\cite{Arndt2}&   144  &     597& - \\ 
$^{139}\rm{Sb}$ &   $93 {^{+14}_{-3}} $~\cite{Arndt2}&     146&     502& - \\ 
\hline
 \multicolumn{5}{c}{ Nuclides  around the  N=126 shell closure}\\
\hline
$^{194}\rm{Re}$ &   $ 1 {^{+0.5}_{-0.5}}$ (s)~\cite{Benlliure2}     &   20.8  (s) & 70.8 (s)& 2.1 (s)  \\  
$^{195}\rm{Re}$ &   $ 6 {^{+1}_{-1}}$ (s)~\cite{Benlliure2}  &   23.9 (s)&  3.3 (s)& 8.5 (s)  \\ 
$^{196}\rm{Re}$ &   $ 3 {^{+1}_{-2}}$    (s)~\cite{Benlliure2}   &  8.8 (s)  & 3.6 (s) &  1.4 (s) \\ 
$^{199}\rm{Os}$ &   $ 5 {^{+4}_{-2}}$ (s)~\cite{Benlliure2} &  13.6(s)&   106.8 (s) & 6.6 (s)  \\ 
$^{200}\rm{Os} $ &   $ 6 {^{+4}_{-3}}$  (s)~\cite{Benlliure2}   & 21.7 (s)&   187.1 (s)& 6.9 (s)    \\  
$^{199}\rm{Ir}$ &   $ 6 {^{+5}_{-4}}$  (s)~\cite{Benlliure2}   &73 (s)&   370.6 (s)& 46.7 (s)   \\  
$^{200}\rm{Ir}$ &   $ 43 {^{+6}_{-5}}$  (s)~\cite{Benlliure2}   &   18.2 (s)   &124.1  (s)& 25.0 (s)    \\ 
$^{201}\rm{Ir}$ &   $ 21 \pm 5$  (s)~\cite{Benlliure2}   &    20.5 (s)&    130.0 (s) & 28.4 (s)  \\  
$^{202}\rm{Ir}$ &   $ 11 {^{+3}_{-3}}$  (s)~\cite{Benlliure2}   & 8.6 (s)&    68.4 (s)& 9.8 (s)  \\ 
$^{203}\rm{Pt}$ &   $ 22  \pm 4$ (s)~\cite{Benlliure2}    & 32.8 (s)&  654.0 (s) & 12.7 (s)  \\   
$^{204}\rm{Pt}$ &   $ 16 {^{+6}_{-5}}$   (s)~\cite{Benlliure2}   & 76.2 (s)&  321.8 (s)  &7.4 (s)   \\
\hline
  & $\sigma_{\rm rms}$  & 0.37& 0.64  & -    \\
 \hline
 \hline 
\caption{Beta-decay half-lives $T_{\beta^{-}}$ of recently 
measured r-process nuclides beyond NUBASE03~\cite{NUBASE03} as produced
by the ANN model, in comparison with experimental values and with results 
from available $pn$QRPA+ffGT~\cite{Moller} and DF3+CQRPA~\cite{Borzov} 
calculations. The overall error measures $\sigma_{\rm rms}$ for ANN and 
$pn$QRPA+ffGT models are  given in the table. Results marked with stars 
are also given in Table IX of Ref.~\cite{Costiris09}.}
\label{TabIeN126}
\end{longtable*}

In Table~\ref{TabIe:Kratz-nn-1026} we present the ANN predictions 
along with calculated $pn$QRPA+\textit{ff}GT~\cite{Moller} values 
for the $\beta^-$ half-lives of nuclides newly identified 
by the PRESPEC collaboration at the GSI FRS/ESR facility.
First we list the half-life values forecast by the two models 
for nuclides in the $N=82$ region (namely $^{122-125}\rm{Rh}$, 
$^{125-128}\rm{Pd}$) that are under study in experiment 
S323~\cite{Dillmann2, Montes} (under analysis).  These predictions are 
also plotted in Figs.~\ref{fig:rh45} and~\ref{fig:pd46}, respectively. 
The waiting-point nucleus $^{128}\rm{Pd}$ acts as a bottleneck of the 
r-process and its decay has implications for the predictions of the 
$\rm{Th}$ and $\rm{U}$ cosmochronometers.  It should be mentioned that 
$^{123-124}\rm{Rh}$ and $^{127-128}\rm{Pd}$ have also been identified at
RIKEN~\cite{Ohnishi}. Second, we list the predicted $T_{\beta^{-}}$ 
values of 14 newly identified nuclides in the $N=126$ region (namely 
$^{205-206}\rm{Ir}$, $^{208-209}\rm{Pt}$, $^{208-211}\rm{Au}$, 
$^{213-214}\rm{Hg}$,$^{213-216}\rm{Tl}$) that have been included 
in experiment S410~\cite{Dillmann2, Tain} (under analysis). 

\begin{table}[h] 
\caption{Beta-decay half-lives $T_{\beta^{-}}$ of nuclides relevant for 
the r-process that are under study at GSI (experiments S323~\cite{Montes} and 
S410~\cite{Tain}), as predicted by the ANN and 
$pn$QRPA+\textit{ff}GT~\cite{Moller} models.}
\begin{ruledtabular}
\begin{tabular}{lll}
& \multicolumn{2}{c}{ $T_{\beta^{-}} (ms)$ }    \\ 
\cline{2-3}  \\ 
Nucleus&  ANN Model~\cite{Costiris09} & $pn$QRPA+\textit{ff}GT~\cite{Moller} \\ 
 \hline
 \multicolumn{3}{c}{Ref.~\cite{Montes}}  \\ 
 \hline
$^{122}\rm{Rh}$&   99.7 & 53.5  \\ 
$^{123}\rm{Rh}$ &  54.9  &   47.9\\ 
$^{124}\rm{Rh}$&    67.4 &  31.0 \\ 
$^{125}\rm{Rh}$ &  46.4  &  25.3\\ 
$^{125}\rm{Pd}$ &  99.4 &  265.4 \\ 
$^{126}\rm{Pd}$ &   92.5 & 221.7  \\ 
$^{127}\rm{Pd}$&    76.6 &  210.1 \\ 
$^{128}\rm{Pd}$&    81.1 &  74.2 \\ 
 \hline
 \multicolumn{3}{c}{Ref.~\cite{Tain}}  \\ 
 \hline
$^{205}\rm{Ir}$&    5.1 (s)& 6.1(s)  \\ 
$^{206}\rm{Ir}$ &     3.4 (s)&  2.9 (s)  \\ 
$^{208}\rm{Pt}$ &    8.8 (s) & 7.1 (s)  \\ 
$^{209}\rm{Pt}$ &    4.2 (s)&   3.7 (s)\\ 
$^{208}\rm{Au}$ &    8.7 (s)&  8.5 (s) \\ 
$^{209}\rm{Au}$  &   9.0 (s) & 7.50 (s)  \\ 
$^{210}\rm{Au}$  &    5.4 (s)&   313.6\\ 
$^{211}\rm{Au}$ &    5.4 (s)&   2.3 (s)\\ 
$^{213}\rm{Hg}$ &    7.0 (s)&  4.7 (s) \\ 
$^{214}\rm{Hg}$ &    9.5 (s)& 3.1 (s)  \\
$^{213}\rm{Tl}$ &         18.2 (s) &  32.4(s) \\ 
$^{214}\rm{Tl}$  &    9.1 (s)&  14.4 (s) \\ 
$^{215}\rm{Tl}$ &    9.4  (s)&  7.8 (s) \\ 
$^{216}\rm{Tl}$  &  5.8 (s)  &  2.8 (s) \\   
\end{tabular}
\end{ruledtabular}
\label{TabIe:Kratz-nn-1026}
\end{table}

In Table~\ref{Table:RIKEN} we present the $\beta^-$ half-life 
predictions of the ANN and $pn$QRPA+\textit{ff}GT~\cite{Moller} models 
for heavy neutron-rich newly produced at RIKEN and GSI, whose half-lives 
have not yet been measured. First, we list the statistical/theoretical 
forecasts for the 41 out of 45 nuclides that are being studied
at RIKEN's RIBF by means of in-flight fission of a $^{238}\rm{U}$ 
beam (from NP0702-RIBF20)~\cite{Ohnishi}.  (Beta half-lives of the 
remaining four nuclides have already been measured; their values can be 
found in Table~\ref{TabIeN126}). Most of these nuclides are considered 
to play a role in the r-process. 
Second, we list the predictions of the two models for the 
half-lives of 55 out of 75 heavy neutron-rich nuclides approaching 
the $A \simeq 195$ region (between $\rm{Yb}$ and $\rm{Fr}$) that have 
been identified at GSI by the RISING collaboration using cold-fragmentation 
reactions of $^{238}\rm{U}$ and $^{208}\rm{Pb}$ projectiles 
of relativistic energies and the fragment separator spectrometer 
FRS~\cite{Benlliure2, Alvarez}. Of the remaining 20 nuclides,
11 already have measured half-lives and 9 are included in the 
S410 experiment; for these we have entered half-life predictions of
both models in Table~\ref{TabIeN126} and Table~\ref{TabIe:Kratz-nn-1026}, 
respectively.  Comparing the predictions of the two models, it
it is found that in most cases the $pn$QRPA+\textit{ff}GT~\cite{Moller} 
model gives longer half-lives than the ANN model.

\begin{table*}[h]
\caption{Beta-decay half-lives $T_{\beta^{-}}$ of nuclides 
relevant for the r-process that have recently been identified at
RIKEN~\cite{Ohnishi} and at GSI~\cite{Benlliure2, Alvarez}, as predicted
by the ANN and $pn$QRPA+\textit{ff}GT~\cite{Moller} models.}

\begin{ruledtabular}
\begin{tabular}{lllllll}
& \multicolumn{2}{c}{ $T_{\beta^{-}} (ms)$ }  &  &&  \multicolumn{2}{c}{ $T_{\beta^{-}} (ms)$ }  \\ 
\cline{2-3}  
 \cline{6-7}\\ 
Nucleus&  ANN Model~\cite{Costiris09} & $pn$QRPA+\textit{ff}GT~\cite{Moller} & & Nucleus &  ANN Model~\cite{Costiris09} & $pn$QRPA+\textit{ff}GT~\cite{Moller} \\ 
 \hline
 \multicolumn{3}{c}{Ref.~\cite{Ohnishi}} & &  \multicolumn{3}{c}{Ref.~\cite{Benlliure2, Alvarez} }   \\ 
 \hline
$^{71}\rm{Mn}$ &   27.1 & 8.1  &&$^{192}\rm{Ta}$  &4.9  (s) &   0.5 (s)  \\
$^{73}\rm{Fe}$ &  38.2  &  56.1 &&$^{193}\rm{Ta}$  & 5.0 (s) &  1.4  (s) \\
$^{74}\rm{Fe}$ &   31.3 &  47.0 &&$^{193}\rm{W}$  &15.2 (s)  & 12.8  (s)  \\
$^{76}\rm{Co}$ &    58 .5&  15.4 &&$^{194}\rm{W}$  & 24.2 (s) &  19.9   (s) \\
$^{79}\rm{Ni}$ &   39.2 &  37.6 &&$^{195}\rm{W}$  & 6.8 (s) & 13.3   (s)  \\
$^{81}\rm{Cu}$ &   52.0 &  80.8 &&$^{197}\rm{Re}$  & 9.4 (s) &   4.5  (s) \\
$^{82}\rm{Cu}$ &   67.2 & 29.3 &&$^{198}\rm{Re}$  & 4.9 (s) & 2.0   (s) \\
$^{84}\rm{Zn}$ &   35.9 &  154.9 && $^{198}\rm{Os}$  & 94.5 (s)  &  601.2  (s) \\
$^{85}\rm{Zn}$ &   31.9 &  19.4 &&$^{201}\rm{Os}$  & 6.6  (s)  &   81.1  (s) \\
$^{87}\rm{Ga}$ &   38.7 &  68.5 &&$^{203}\rm{Ir}$  & 9.0  (s) & 34.9 (s)  \\
 
$^{90}\rm{Ge}$ &    20.6&  43.0 &&$^{205}\rm{Pt}$  & 12.8  (s) &  20.6 (s)  \\
$^{95}\rm{Se}$ &    27.4&  25.8 &&$^{206}\rm{Au}$  & 17.0 (s) & 21.3  (s)  \\
$^{98}\rm{Br}$ &    44.4&  29.0 &&$^{207}\rm{Au}$  & 18.8 (s) &22.3   (s)  \\
$^{101}\rm{Kr}$ &    16.1&  14.3 &&$^{211}\rm{Hg}$  & 12.6  (s) & 14.9   \\ 
$^{103}\rm{Rb}$ &    18.2&  16.5 && $^{212}\rm{Hg}$  & 21.1  (s) &    10.5\\ 
$^{106}\rm{Sr}$ &    10.7  &  39.7 &&  $^{215}\rm{Hg}$  & 4.5  (s) &  0.2   (s) \\ 
$^{107}\rm{Sr}$ &    10.9&  28.0 && $^{217}\rm{Tl}$  &5.7  (s) &  1.3  (s)  \\
$^{109}\rm{Y}$ &    12.9&  18.4 &&$^{215}\rm{Pb}$  &27.7 (s)  &282.5     (s) \\
$^{111}\rm{Zr}$ &  15.8  &  50.7 &&$^{216}\rm{Pb}$  &67.0  (s) &   852.2  (s) \\
$^{112}\rm{Zr}$ & 9.1  &  42.7 &&$^{217}\rm{Pb}$  & 12.9 (s) &   104.9  (s) \\

  $^{114}\rm{Nb}$   &    27.2&  21.0 &&$^{218}\rm{Pb}$  &22.4 (s)  &   66.3  (s) \\
  $^{115}\rm{Nb}$   &    11.1&   15.5 &&$^{219}\rm{Pb}$  &7.5  (s) &  1.2  (s)  \\
  $^{116}\rm{Mo}$   &    15.6& 50.6  &&$^{220}\rm{Pb}$  &10.4  (s) & 7.0   (s)  \\
  $^{117}\rm{Mo}$   &    14.7&  42.9 &&$^{219}\rm{Bi}$  & 18.5 (s) & 26.5   (s)  \\
  $^{119}\rm{Tc}$   &    20.6&  25.1 &&$^{220}\rm{Bi}$  & 9.6 (s) & 5.2   (s)  \\
 $^{120}\rm{Tc}$   &    28.0&  21.8 &&$^{221}\rm{Bi}$  & 10.0 (s) &  9.7   (s) \\
 $^{121}\rm{Ru}$   &    31.8&   88.3 &&$^{222}\rm{Bi}$  &6.3  (s) &  2.0   (s) \\
 $^{122}\rm{Ru}$   &   26.4 &  68.6  &&$^{223}\rm{Bi}$  & 6.2 (s) &   3.5  (s) \\
  $^{123}\rm{Ru}$   &    24.5&  50.6  &&$^{224}\rm{Bi}$  & 4.5 (s) &  2.4  (s)  \\
  $^{124}\rm{Ru}$   &   24.1 &  45.0 &&$^{221}\rm{Po}$  & 27.8 (s) & 3.2   (s)  \\
  
  $^{123}\rm{Rh}$   &    54.9&   47.9 &&$^{222}\rm{Po}$  &70.9 (s)  &  27.3  (s)  \\
$^{124}\rm{Rh}$   &    67.4&   31.0 &&$^{223}\rm{Po}$  &13.7  (s) &  4.4  (s)  \\
  $^{125}\rm{Rh}$   &   46.4 &   25.3 &&$^{224}\rm{Po}$  &24.9  (s) &   11.1  (s) \\
$^{126}\rm{Rh}$   &   57.7 &   26.0 &&$^{225}\rm{Po}$  & 8.1 (s) &   5.7  (s) \\
  $^{127}\rm{Pd}$   &    76.6&  210.1 &&$^{226}\rm{Po}$  & 11.8 (s) &3.0    (s)  \\
 $^{128}\rm{Pd}$   &    81.1&   74.2 &&$^{227}\rm{Po}$  & 5.4 (s) & 2.4 (s)    \\
 $^{140}\rm{Sb}$   &    82.0&  359.1 &&$^{224}\rm{At}$  & 17.7 (s) & 4.1   (s)  \\
 $^{143}\rm{Te}$   &    112.3&   63.8 &&$^{225}\rm{At}$  & 19.5 (s) &  5.1  (s)  \\
 $^{145}\rm{I}$  &    147.8&  55.2 &&$^{226}\rm{At}$  &10.4  (s) &  3.3  (s)  \\
$^{148}\rm{Xe}$  &  159.6  &  116.8 &&$^{227}\rm{At}$  & 10.8 (s) & 8.1   (s)  \\

 $^{152}\rm{Ba}$ &    261.2&  191.2 &&$^{228}\rm{At}$  &6.9  (s) &  2.1  (s)  \\
 \cline {1-3}
 \multicolumn{3}{c}{Ref.~\cite{Benlliure2, Alvarez} }  &&$^{229}\rm{At}$  & 6.9 (s) &   1.7  (s) \\
 \cline {1-3}
$^{183}\rm{Yb}$  & 7.9 (s)   & 31.8   (s)  &&$^{230}\rm{At}$  & 4.9 (s) &  0.7  (s)  \\
$^{184}\rm{Yb}$  & 9.4  (s) &  24.9   (s) &&$^{229}\rm{Rn}$  &14.8  (s) &  21.8  (s)  \\
$^{186}\rm{Lu}$  & 5.0 (s) &  2.1  (s)  &&$^{230}\rm{Rn}$  & 28.7 (s) &  12.6  (s)  \\
$^{187}\rm{Lu}$  & 5.2 (s) & 1.4 (s)    &&$^{231}\rm{Rn}$  &8.9  (s) &  3.9  (s)  \\
$^{189}\rm{Hf}$  &7.2 (s)  &  1.2  (s)  &&$^{232}\rm{Rn}$  & 13.7 (s) &  4.5 (s) \\
$^{190}\rm{Hf}$  & 8.8 (s) &  4.0 (s)   &&$^{233}\rm{Fr}$  & 11.9  (s) & 9.4 (s)    \\
$^{191}\rm{Ta}$  &10.2  (s) &   1.6 (s) &&  &  &    \\
\end{tabular}
\end{ruledtabular}
\label{Table:RIKEN}
\end{table*}

\section{\label{sec:level-4}Conclusions}

The study of r-process nucleosynthesis is currently among the most
vibrant areas of research in nuclear astrophysics.  It is also one
of the most challenging, as key issues remain open with respect
to both the nature of the astrophysical setting and the nuclear
physics input, which entails knowledge of nuclear behavior far
from the valley of stability.  This paper has been concerned
with the latter aspect of r-process physics, as manifested
in the $\beta^{-}$-decay half-lives of nuclei that are believed
to play important roles in the r-process under different astrophysical
scenarios.  We have explored, in quantitative detail, the predictive
capacity of a statistical global model of $\beta^{-}$-decay systematics
referred to as the standard ANN model~\cite{Costiris09}.
This model is an artificial neural network (ANN) of feedforward 
multilayer perceptron architecture that has been trained to map a 
given nuclidic input ($Z,N$) to an output encoding the corresponding 
$\beta^-$ half-life $T_{\beta^{-}}$ of the nuclear ground state.  The 
training is performed by advanced algorithms for supervised learning based 
on a subset of data selected at random from NUBASE03, which tabulates
$\beta^{-}$ half-lives measured before November 2003.  Good quality of 
the ANN model is substantiated in its values for the performance measure
provided by the root-mean-square error $\sigma_{\rm rms}$, namely
0.53, 0.60, and 0.65 on the training, validation, and test sets
of data selected from NUBASE03.

Predictive power (``extrapability'') is properly measured by performance
on test nuclei {\it outside} the training and validation sets, since
such nuclides have no influence on the development of the statistical
model. Of special interest in the present context is the predictive
performance on r-process nuclides studied at rare-isotope-beam
facilities since the publication of NUBASE03.  It should be emphasized
that the half-lives produced by the standard ANN model that appear
in Tables II-IV are all {\it true predictions} in this sense, because
they only involve target nuclides {\it beyond} NUBASE03.

In addition to comparing the half-lives produced by the data-driven
ANN model with available experimental $\beta^-$ half-lives, it is
of obvious importance to compare both the behavior of its outputs 
and the quality of its performance with that of established
theory-thick nuclear models. With respect to behavior, the general 
trend of such a comparison is exemplified in the finding
that the half-live values given by the standard ANN model are
lower than those obtained in the $pn$QRPA+\textit{ff}GT
calculations of M\"oller et al.~\cite{Moller} and closer to those from 
the DF3+CQRPA~\cite{Borzov} calculations.  The odd-even staggering 
effect is less pronounced than that found in Ref.~\cite{Moller}.
The relative values of the root-mean-square error $\sigma_{\rm rms}$ 
characterizing the ANN and $pn$QRPA+\textit{ff}GT
half-life predictions for the set of all nuclides with half-lives 
measured subsequent to NUBASE03 (cf.\ Table~\ref{TabIeN126}), 
respectively 0.37 and 0.64, are symptomatic of good extrapability 
for the ANN model.  A number of additional tests of the reliability
of the standard ANN model have been reported in Ref.~\cite{Costiris09}. 

In conclusion, the available evidence indicates that the theory-thin
ANN statistical global model of $\beta^-$ half-lives explored in the
present work can provide a robust tool for generating nuclear input to
r-process clock and matter-flow studies, and therefore serves to 
{\it complement} conventional theory-thick nuclear modeling.  Naturally, 
the traditional models will continue to be superior in revealing
the underlying physics responsible for the values taken by the
targeted nuclear observables.

A promising strategy for further improvement of statistical global
modeling of $\beta^-$-decay systematics lies in the development
of a hybrid ANN model that targets the set of {\it differences}
between true or experimental half-life values and the corresponding 
values generated by a conventional theory-thick model (e.g., 
that of M\"oller et al.~\cite{Moller}.  Such an approach
has had demonstrable success in global modeling of atomic 
masses~\cite{Athanassopoulos}.  Advances in predictive accuracy
may also be sought through committee-machine strategies,
in which different ANNs are built to process given input patterns and
decide on the best output by a voting process.  

\section{\label{sec:level-5}Acknowledgments}

This research has been supported in part by the University of Athens 
under Grant No. 70/4/3309. The authors wish to thank I. N. Borzov, 
T. Marketin, P. M\"{o}ller, and P. Ring for supplying us with 
theoretical data and for helpful discussions and/or correspondence.

\newpage  
\bibliography{paperbib}

\end{document}